\newif\ifcameraready

\camerareadytrue

\documentclass[sigconf]{acmart}

\usepackage{algorithm}

\usepackage{algorithmic} 

\usepackage{amsmath}
\usepackage{graphicx}
\usepackage{multirow}
\usepackage{booktabs}
\usepackage{bm}
\usepackage{xcolor}
\usepackage{subcaption}
\usepackage{siunitx}
\usepackage{enumitem}

\usepackage{xspace}

\usepackage[]{hyperref}

\newcommand{\PRE}{\texttt{PRE}}
\newcommand{\ACT}{\texttt{ACT}}

\newcommand{\WRITE}{\texttt{WRITE}}

\newcommand{\myAND}{\texttt{AND}}
\newcommand{\myOR}{\texttt{OR}}
\newcommand{\myNOT}{\texttt{NOT}}

\newcommand{\RowCopy}{\texttt{RowCopy}}
\newcommand{\MAJ}[1]{\texttt{MAJ#1}}
\newcommand{\Frac}{\texttt{Frac}}

\newcommand{\B}{$B$}
\newcommand{\Y}{$Y$}
\newcommand{\smallA}{$a$}

\newcommand{\bcircled}[1]{%
  \begingroup
  \setlength{\unitlength}{1em}%
  \begin{picture}(1.2,1)
    \put(0.6,0.35){\circle*{1.1}}%
    \put(0.6,0.35){\makebox(0,0){\color{white}\small\bfseries #1}}%
  \end{picture}%
  \endgroup
}

\newcommand{\wcircled}[1]{%
  \begingroup
  \setlength{\unitlength}{1em}%
  \begin{picture}(1.2,1)
    \put(0.6,0.35){\circle{1.1}}%
    \put(0.6,0.35){\makebox(0,0){\small\bfseries #1}}%
  \end{picture}%
  \endgroup
}

\definecolor{gfored}{rgb}{0.580, 0.050, 0.211}
\definecolor{ao}{rgb}{0.007, 0.520, 0.867}
\definecolor{moegi}{rgb}{0.357, 0.537, 0.188}
\definecolor{jl}{rgb}{1.0, 0.2, 0.8}
\definecolor{brown(web)}{rgb}{0.65, 0.16, 0.16}
\definecolor{bisque}{rgb}{1.0, 0.89, 0.77}
\definecolor{nbs}{rgb}{0.88, 0.07, 0.37}
\definecolor{yt}{rgb}{0.58, 0.44, 0.86}
\definecolor{iy}{rgb}{0.0, 0.56, 0.041}
\definecolor{mel}{rgb}{0.9, 0.55, 0.31}
\definecolor{ouscolor}{rgb}{0.0, 0.2, 0.4}
\definecolor{dt}{rgb}{0.5, 0.1, 1.0}

\newcommand{\concern}[1]{\colorbox{yellow!40}{\hyperref[ref:#1]{{#1}}}}

\newcommand{\cref}[1]{\hyperref[ref:#1]{#1}}

\ifcameraready

    \newcommand{\atbcr}[2]{\ifnum#1>-1\textcolor{black}{#2}\else{#2}\fi}
    \newcommand{\ieycr}[2]{\ifnum#1>-1\textcolor{black}{#2}\else{#2}\fi}
    \newcommand{\omcr}[2]{\ifnum#1>-1\textcolor{black}{#2}\else{#2}\fi}
    \newcommand{\dtcr}[2]{\ifnum#1>-1\textcolor{black}{#2}\else{#2}\fi}
    \newcommand{\omcrcomment}[1]{}
    \newcommand{\crdiscussion}[2]{}{}

    \newcommand{\ominline}[1]{}
    \newcommand{\ieycrcomment}[1]{}
    \newcommand{\atbcrcomment}[1]{}
    \newcommand{\agycrcomment}[1]{}
    \newcommand{\dtcrcomment}[1]{}
    \newcommand{\ieyinline}[1]{}
    \newcommand{\dtinline}[1]{}
\else
\usepackage[colorinlistoftodos,prependcaption,textsize=scriptsize]{todonotes}
    \paperwidth=\dimexpr \paperwidth + 4cm\relax
    \oddsidemargin=\dimexpr\oddsidemargin + 2cm\relax
    \evensidemargin=\dimexpr\evensidemargin + 2cm\relax
    \marginparwidth=\dimexpr \marginparwidth + 2cm\relax
    
    \newcommand{\atbcr}[2]{\ifnum#1=\value{version}\textcolor{ao}{#2}\else{#2}\fi}

    \newcommand{\ieycr}[2]{\ifnum#1=\value{version}\textcolor{blue}{#2}\else{#2}\fi}

    \newcommand{\dtcr}[2]{\ifnum#1=\value{version}\textcolor{dt}{#2}\else{#2}\fi}

    \newcommand{\ieycrcomment}[1]{\todo[linecolor=orange, bordercolor=orange, backgroundcolor=white]{\textcolor{iy}{\textbf{@Ismail:} #1}}}
    \newcommand{\ieyinline}[1]{\\\textcolor{iy}{\textbf{@Ismail:} #1}}

    \newcommand{\dtcrcomment}[1]{\todo[linecolor=orange, bordercolor=orange, backgroundcolor=white]{\textcolor{dt}{\textbf{@Daichi:} #1}}}
    \newcommand{\dtinline}[1]{\\\textcolor{dt}{\textbf{@Daichi:} #1}}

    \newcommand{\atbcrcomment}[1]{\todo[linecolor=brown, bordercolor=brown, backgroundcolor=white]{\textcolor{ao}{\textbf{@Atb:} #1}}}

    \newcommand{\agycrcomment}[1]{\todo[size=\scriptsize, linecolor=orange, bordercolor=orange, backgroundcolor=white]{\textcolor{gfored}{\textbf{@gy:} #1}}}
    
    \newcommand{\crdiscussion}[3]{\omcrcomment{#1\\\textcolor{dt}{\textbf{@Daichi:}#2}\\\textcolor{blue}{\textbf{@Ismail:}#3}}}
    \newcommand{\omcr}[2]{\ifnum#1=\value{version}\textcolor{red}{#2}\else{#2}\fi}

    \newcommand{\omcrcomment}[1]{\todo[linecolor=orange, bordercolor=orange, backgroundcolor=white]{\textcolor{red}{\textbf{@Onur:} #1}}}
    \newcommand{\ominline}[1]{\\\textcolor{red}{\textbf{@Onur:} #1}}

\fi

\AtBeginDocument{%
  }

\copyrightyear{2026}
\acmYear{2026}
\setcopyright{cc}
\setcctype{by}
\acmConference[ICS '26]{2026 International Conference on Supercomputing}{July 06--09, 2026}{Belfast, United Kingdom}
\acmBooktitle{2026 International Conference on Supercomputing (ICS '26), July 06--09, 2026, Belfast, United Kingdom}
\acmDOI{10.1145/3797905.3800550}
\acmISBN{979-8-4007-2522-7/2026/07}

\begin{document}

\title[Clutch: High Performance Vector-Scalar Comparison using DRAM via Chunked Temporal Coding]{Clutch: High Performance Vector-Scalar Comparison \\using DRAM via Chunked Temporal Coding}

\author{Daichi Tokuda}
\affiliation{%
  \institution{The University of Tokyo}
  \city{Tokyo}
  \country{Japan}}
\email{tokuda-daichi@is.s.u-tokyo.ac.jp}
\orcid{0009-0006-9328-3805}

\author{Tatsuya Kubo}
\affiliation{%
  \institution{The University of Tokyo / RIKEN}
  \city{Tokyo}
  \country{Japan}}
\email{tatsuya.kubo@is.s.u-tokyo.ac.jp}
\orcid{0009-0005-1523-3063}

\author{Ismail Emir Yuksel}
\affiliation{%
  \institution{ETH Zurich}
  \city{Zurich}
  \country{Switzerland}}
\email{ismailemryksel@gmail.com}
\orcid{0000-0003-3310-4423}

\author{Ataberk Olgun}
\affiliation{%
  \institution{ETH Zurich}
  \city{Zurich}
  \country{Switzerland}}
\email{olgunataberk@gmail.com}
\orcid{0000-0001-5333-5726}

\author{Haocong Luo}
\affiliation{%
  \institution{ETH Zurich}
  \city{Zurich}
  \country{Switzerland}}
\email{richardluo723@gmail.com}
\orcid{0009-0009-0849-8724}

\author{Tomoya Nagatani}
\affiliation{%
  \institution{The University of Tokyo}
  \city{Tokyo}
  \country{Japan}}
\email{t-nagatani@is.s.u-tokyo.ac.jp}
\orcid{0009-0000-0282-743X}

\author{Geraldo F. Oliveira}
\affiliation{%
  \institution{ETH Zurich}
  \city{Zurich}
  \country{Switzerland}}
\email{geraldofojunior@gmail.com}
\orcid{0000-0003-1557-4819}

\author{Abdullah Giray Ya\u{g}l{\i}k\c{c}{\i}}
\affiliation{%
  \institution{CISPA}
  \city{Saarbr\"{u}cken}
  \country{Germany}}
\email{agirayyaglikci@gmail.com}
\orcid{0000-0002-9333-6077}

\author{Mohammad Sadrosadati}
\affiliation{%
  \institution{ETH Zurich}
  \city{Zurich}
  \country{Switzerland}}
\email{m.sadr89@gmail.com}
\orcid{0000-0002-4029-0175}

\author{Onur Mutlu}
\affiliation{%
  \institution{ETH Zurich}
  \city{Zurich}
  \country{Switzerland}}
\email{omutlu@gmail.com}
\orcid{0000-0002-0075-2312}

\author{Shinya Takamaeda-Yamazaki}
\affiliation{%
  \institution{The University of Tokyo / RIKEN}
  \city{Tokyo}
  \country{Japan}}
\email{shinya@is.s.u-tokyo.ac.jp}
\orcid{0000-0003-3441-1695}

\renewcommand{\shortauthors}{Daichi Tokuda et al.}

\begin{abstract}
Vector–scalar comparison is a fundamental computation primitive that compares each element in a vector against a single scalar value. It is widely used in a broad range of data-intensive workloads from databases to machine learning. Due to its low computational intensity, the execution of this operation tends to be memory-bound, especially for large vectors, thereby limiting the utilization of compute resources. Processing-using-DRAM (PuD) is an emerging computing paradigm that performs massively parallel bitwise operations directly within the DRAM array, alleviating off-chip data movement.
Unfortunately, no prior work proposes an efficient PuD-based solution tailored to vector–scalar comparisons.
Existing PuD-based approaches require many DRAM commands because the comparison’s algorithmic complexity grows with operand \dtcr{2}{bit-width} in the bit-serial execution model, which is inherently induced by current PuD architectures. As a result, this command overhead becomes the dominant performance bottleneck, limiting application-level \dtcr{2}{speedup}.

We propose Clutch, a novel data representation and comparison algorithm for accelerating vector–scalar comparisons in PuD systems with high efficiency and scalability. Our key idea is twofold. First, to reduce the number of DRAM commands required for comparison, Clutch adopts temporal coding for vectors, where each value is encoded as a sequence of leading ones. This enables lookup-based comparisons, where comparing against a scalar input simply involves accessing the corresponding DRAM row. Second, Clutch leverages our key insight that a divide-and-conquer approach enables scalable lookup-based comparisons without incurring a prohibitive memory footprint at high \dtcr{2}{bit-precision}. \dtcr{1}{Specifically, Clutch partitions the operand into multiple multi-bit chunks which can be compared independently using compact lookup tables}, \dtcr{2}{and merges per-chunk results through a procedure designed to execute efficiently on PuD.}
Clutch provides a flexible \dtcr{3}{tradeoff} between throughput and memory usage by adjusting chunk count.

Experimental results on two applications, predicate evaluation and decision tree inference,
demonstrate that Clutch improves end-to-end application throughput (and energy efficiency) by an average of 12$\times$ (69$\times$) over highly optimized CPU and GPU execution and 2.9$\times$ (3.0$\times$) over \dtcr{3}{the state-of-the-art bit-serial PuD implementation}.
Notably, we present, to our knowledge, the first mapping of decision tree inference to PuD execution, extending PuD to a new application domain. Our results demonstrate that DRAM can serve as a high-performance \dtcr{3}{and energy-efficient} computing \dtcr{3}{substrate} for comparison-intensive workloads.
\end{abstract}




\maketitle

\section{Introduction}
Vector-scalar comparison, which compares every element in a vector with a single scalar value, is a fundamental primitive in data-intensive workloads. It is widely used in predicate evaluation for in-memory database query processing~\cite{willhalm2009simd, graefe2011modern, li2013bitweaving, farber2012sap, grund2010hyrise, idreos2012monetdb, kemper2012hyper, lahiri2015oracle}, thresholding and masking operations in scientific computing and image processing~\cite{ergin2017dynamic, zhong2022using, santitissadeekorn2020approximate, abdusalomov2020automatic, Li2021_HTMaskRCNN}, and decision tree ensemble inference in machine learning~\cite{chen2016xgboost, ke2017lightgbm, prokhorenkova2018catboost, hancock2020catboost}.

Efficient vector–scalar comparison is critical in many of these data-intensive applications because the performance of this step often dominates overall execution~\cite{khataei2025treelut,willhalm2009simd,li2013bitweaving}. Our profiling \dtcr{3}{results on real CPU systems} show that comparisons account for up to \dtcr{3}{1)} 96\% of execution time in query processing workloads~\cite{willhalm2009simd, graefe2011modern, li2013bitweaving, farber2012sap, grund2010hyrise, idreos2012monetdb, kemper2012hyper, lahiri2015oracle} and \dtcr{3}{2)} 81\% \dtcr{3}{of execution time} in Gradient Boosting Decision Tree (GBDT) inference~\cite{prokhorenkova2018catboost,hancock2020catboost} \dtcr{2}{(see \S\ref{sec:mot:prof} for the detailed methodology and configuration)}. However, due to its \dtcr{3}{very} low computational intensity, the performance of vector–scalar comparison is \dtcr{2}{typically} bottlenecked by main-memory bandwidth, leaving compute resources underutilized on large-scale data.

Processing-using-Memory (PuM)~\cite{aga2017computecaches, ali2019inmemoryaddition, angizi2019graphide, besta2021sisa, bostanci2022drstrange, chi2016prime, deng2018dracc, eckert2018neuralcache, ferreira2021plutoarxiv, ferreira2022pluto, fujiki2018inmemorydataparallel, fujiki2019dualitycache, gao2019computedram, he2020sparsebdnet, imani2019floatpim, kim2018dramlatencypuf, kim2019drange, li2018scope, li2017drisa, li2016pinatubo, olgun2022pidram, olgun2021quac, orosa2021codic, park2022flashcosmos, rezaei2020nom, seshadri2015fast, seshadri2013rowclone, seshadri2018rowclonearxiv, seshadri2016buddy, seshadri2017ambit, seshadri2016processing, seshadri2017simpleoperations, seshadri2019dram, shafiee2016isaac, sharad2013ultralowpowerassociative, song2017pipelayer, song2018graphr, subramaniyan2017parallelautomata, truong2021racer, truong2022raceradapt, xin2020elp2im, zha2020hyperap, mutlu2024memory, mutlu2025memory} has emerged as a promising approach to alleviate the processor-memory data movement bottleneck by performing computation directly within memory arrays using the intrinsic operational principles of memory cells. Among various PuM technologies, \dtcr{2}{Processing-using-DRAM (PuD)~\cite{kim2012case, seshadri2013rowclone, seshadri2015fast, seshadri2016buddy, seshadri2016processing, li2017drisa, deng2018dracc, seshadri2019dram,deng2019lacc, wang2020figaro, xin2020elp2im, hajinazar2021simdram, ferreira2022pluto, wu2022dram, deng2023dram, oliveira2024mimdram, gao2019computedram, gao2022fracdram, yuksel2024simultaneous, kubo2025pudtune, kubo2025mvdram, chang2016low, shivanandamurthy2021atria, afifi2024artemis, jahshan2024majork, mutlu2024memory, liu2025optipim, de2026count2multiply, garzon2026cadm, mutlu2025memory} leverages the analog operational principles of DRAM \dtcr{3}{circuitry} and the massive internal parallelism of each DRAM bank, enabling massively parallel computations within DRAM}. A key design principle in conventional PuD architectures~\cite{seshadri2017ambit, gao2019computedram, hajinazar2021simdram, oliveira2024mimdram, oliveira2025proteus, kubo2025pudtune, kubo2025mvdram} is to perform arithmetic in a bit-serial manner within each column through repeated PuD operations (i.e., bulk data copy and bitwise operation), each implemented using dedicated DRAM command sequences.

Our PuD performance analysis reveals that PuD is particularly effective for vector–scalar \dtcr{3}{comparison} in terms of data movement efficiency (see \S\ref{sec:mot:ana}). In a processor-centric \dtcr{3}{system}, the entire input vector must be read from memory and transferred to the processor, where each element \dtcr{3}{of the vector} is compared against the \dtcr{3}{same} scalar \dtcr{3}{value}. In contrast, PuD performs the comparison directly within DRAM, reducing data movement to only a 1-bit-per-element result bitmap. PuD also enables high-throughput subsequent processing with reduced data movement. In real-world workloads, comparison results often need to be combined with other bitmaps \dtcr{3}{via} reduction operations such as \myAND/\myOR{} (see \S\ref{sec:app}). PuD can perform such reductions directly in DRAM at high throughput \dtcr{3}{using bulk bitwise operations}, eliminating the need to transfer intermediate bitmaps back to the processor.

\dtcr{2}{However, PuD-based execution \dtcr{3}{sometimes} shows limited performance improvements at the application level, and can even underperform processor-based execution under certain configurations (see Figure~\ref{fig:gbdt_thr} and Figure~\ref{fig:ts_q5_gpu})}.
This occurs because conventional PuD approaches~\cite{seshadri2017ambit, gao2019computedram, hajinazar2021simdram, oliveira2024mimdram, oliveira2025proteus, liu2025optipim, kubo2025pudtune, kubo2025mvdram} rely on \dtcr{3}{\emph{bit-serial execution}}, \dtcr{3}{in which the comparison's algorithmic complexity grows with the operand bit-width, leading to a large number of PuD operations.}
Despite reduced data movement, the performance bottleneck shifts from off-chip data movement to the latency of PuD operations required for in-DRAM comparisons, \dtcr{3}{limiting application-level performance}. \dtcr{3}{As such}, reducing the number of PuD operations is a key challenge for unlocking the full potential of PuD for accelerating comparison-intensive applications.

We present \textit{\textbf{Clutch}} (\underline{C}omparison Algorithm using \underline{L}ook\underline{u}p \underline{T}able with \underline{Ch}unked Temporal Coding), a new data representation and algorithm designed for PuD execution to enable efficient and scalable vector–scalar comparisons. Our key idea is \dtcr{5}{twofold}. 
First, to reduce the number of PuD operations required for comparison, Clutch operates on vectors encoded with \dtcr{3}{\emph{temporal coding}}~\cite{madhavan2014race,wu2020ugemm} instead of \dtcr{3}{the binary representation}. In this scheme, a value $v$ is represented as a sequence of $v$ leading ones followed by zeros (e.g., a 3-bit value of 3 is encoded as 1110000). \dtcr{3}{A key property of temporal coding is that the $i$-th bit of the encoded value $v$ equals the truth value of $i < v$. As a result, when vector elements are stored in temporal coding across the columns of a DRAM subarray, each row directly contains the output bitmap of a vector–scalar comparison for the corresponding scalar value. The host processor can therefore execute a vector–scalar comparison with a single \RowCopy{}, producing the result within DRAM} and significantly reducing the number of PuD operations compared to the bit-serial approach.

Second, to support high bit-precision \dtcr{3}{(e.g., 16-bit and 32-bit)} in a scalable and memory-efficient manner, Clutch introduces a divide-and-conquer approach based on our key insight that full-width comparisons can be decomposed into comparisons on smaller bit chunks. \dtcr{2}{Specifically, Clutch partitions the representation of each operand into multiple multi-bit chunks.} Each chunk is independently compared \dtcr{3}{to} the corresponding portion of the scalar \dtcr{3}{value} by referencing a compact lookup table encoded with temporal coding. \dtcr{3}{These per-chunk results are then merged through a procedure that propagates carry information across chunks, optimized for PuD execution.}
This chunk-wise design significantly reduces \dtcr{3}{the number of DRAM rows used} compared to encoding the full bit-width as a single lookup table. By adjusting the number of chunks, Clutch \dtcr{5}{provides} a flexible tradeoff between comparison throughput and memory footprint.

We demonstrate Clutch's effectiveness on two comparison-
intensive applications: predicate evaluation for in-memory databases~\cite{willhalm2009simd, graefe2011modern, li2013bitweaving, farber2012sap, grund2010hyrise, idreos2012monetdb, kemper2012hyper, lahiri2015oracle} and \dtcr{3}{Gradient Boosting Decision Tree (GBDT)} inference~\cite{chen2016xgboost, ke2017lightgbm, prokhorenkova2018catboost, hancock2020catboost}. For GBDT, we propose a novel mapping of the inference process onto PuD, based on our observation that tree traversal can be reformulated as a sequence of vector–scalar comparisons followed by mask operations, which Clutch can directly accelerate.
We conduct detailed end-to-end performance evaluations on both applications using two PuD architectures: one requiring no DRAM modifications~\cite{kubo2025mvdram, kubo2025pudtune, jahshan2024majork, garzon2026cadm} and SIMDRAM~\cite{hajinazar2021simdram, seshadri2017ambit, liu2025optipim, de2026count2multiply}, \dtcr{3}{which modifies DRAM circuitry to natively support bulk bitwise \myNOT{} operations}.

The main contributions of this work are as follows:
\begin{itemize}
\item 
Our application profiling and \dtcr{5}{Processing-using-DRAM (PuD)} performance analysis reveal that reducing the number of PuD operations required for comparisons is critical to accelerating \dtcr{3}{comparison-intensive} applications.
\item
\dtcr{3}{We present Clutch, a new PuD-oriented comparison algorithm and data representation, \dtcr{5}{which} uses a lookup-table-based method with temporal coding to significantly reduce the number of PuD operations required for comparison.}
\dtcr{3}{Clutch adopts a divide-and-conquer approach that enables a flexible tradeoff between throughput and memory footprint by adjusting the number of chunks.}
\item 
We apply Clutch to two applications: predicate evaluation and \dtcr{3}{Gradient Boosting Decision Tree (GBDT)} inference. To our knowledge, this is the first work that applies PuD-based acceleration to GBDT. On average, Clutch improves throughput (and energy efficiency) by $12\times$ ($69\times$) over optimized processor execution and by $2.9\times$ ($3.0\times$) over the state-of-the-art bit-serial PuD approach.
\item 
Our results demonstrate that PuD can provide substantial application-level \dtcr{3}{speedup} when paired with an algorithm \dtcr{3}{and data representation} that \dtcr{5}{together} minimize PuD operation count, highlighting DRAM's viability as a high-performance \dtcr{3}{and energy-efficient} computing \dtcr{3}{substrate}.
\end{itemize}



\section{Background}\label{sec:bac}
\subsection {\dtcr{3}{Vector-Scalar Comparison}}
Vector-scalar comparison is a fundamental and widely used operation across various data processing pipelines. It typically involves comparing each element in a vector against a \dtcr{3}{single scalar value}, followed by filtering, masking, or other conditional processing based on the results. Such vector-scalar comparisons are central to predicate evaluation workloads, such as query processing in in-memory databases and thresholding in scientific and image processing~\cite{ergin2017dynamic, zhong2022using, santitissadeekorn2020approximate, abdusalomov2020automatic, Li2021_HTMaskRCNN}. In the machine learning domain, this operation also plays a dominant role in the inference of Gradient Boosting Decision Tree (GBDT)~\cite{chen2016xgboost, ke2017lightgbm, prokhorenkova2018catboost, hancock2020catboost}.

These applications play a foundational role in modern data-centric systems that support real-world infrastructure. Optimizing predicate evaluation boosts query performance in in-memory databases~\cite{willhalm2009simd, graefe2011modern, li2013bitweaving, farber2012sap, grund2010hyrise, idreos2012monetdb, kemper2012hyper, lahiri2015oracle}. GBDT, with its low computational cost and state-of-the-art accuracy on tabular data, enables fast, low-overhead decision-making in edge AI systems with limited resources. More detailed descriptions of these applications are provided in\dtcr{3}{~\S\ref{sec:app:gbdt} and~\S\ref{sec:app:que}}.

\subsection{DRAM Organization and Operation}
\begin{figure}[htbp]
    \centering
    \includegraphics[width=\columnwidth]{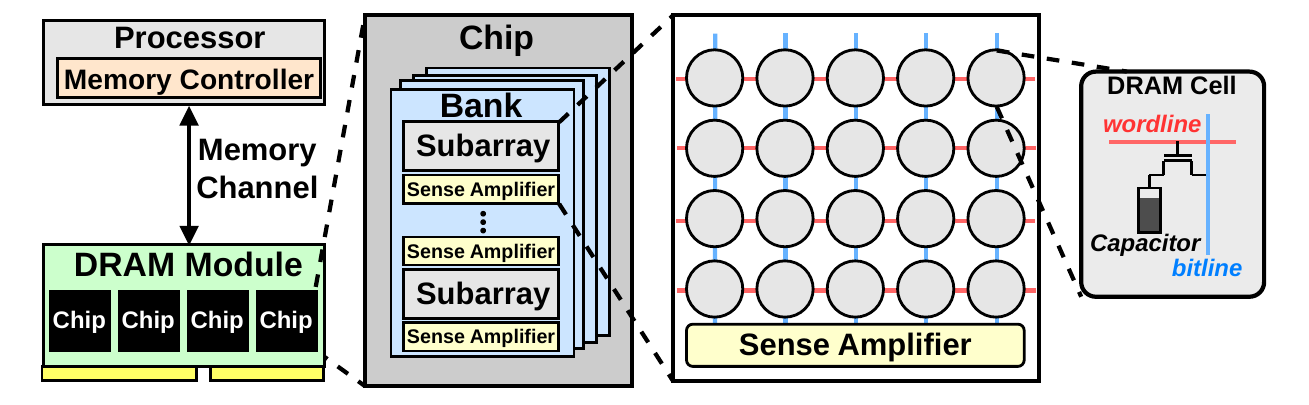}
    \caption{DRAM Organization.}
    \vspace{-0.5em}
    \label{fig:dram}
\end{figure}

Dynamic Random Access Memory (DRAM) is organized in a hierarchical structure consisting of channels, ranks, chips, banks, and subarrays of memory cells (Figure~\ref{fig:dram}). Each cell consists of a single transistor paired with a capacitor. Each cell stores one bit of data, based on the charge \dtcr{5}{level} held in the capacitor. Within each subarray, cells form a two‑dimensional grid of rows (wordlines) and columns (bitlines). The memory controller integrated in the CPU die generates a sequence of DRAM commands to access data in DRAM. The \ACT{} command opens a specific row and copies its data \dtcr{3}{into} the row buffer. The \PRE{} command closes the active row. \dtcr{5}{These commands operate on all columns in a row (e.g., 64K columns per bank in DDR4), and multiple operations can be performed concurrently across different banks (e.g., 16 banks).}

\subsection{Processing-using-DRAM}\label{sec:bac:pud}
\begin{figure}[htbp]
    \centering
    \includegraphics[width=0.85\columnwidth]{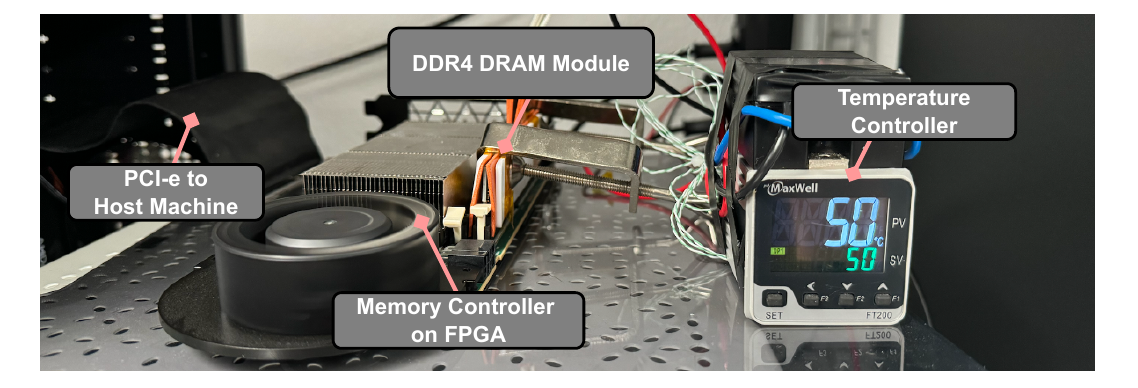}
    \caption{\dtcr{2}{Our FPGA-based PuD testing infrastructure (DRAM Bender~\cite{olgun2023dram}) with DDR4 modules.}}
    \label{fig:fpga}
\end{figure}

Processing-using-DRAM (PuD)~\cite{kim2012case, seshadri2013rowclone, seshadri2015fast, seshadri2016buddy, seshadri2016processing, li2017drisa, deng2018dracc, seshadri2019dram,deng2019lacc, wang2020figaro, xin2020elp2im, hajinazar2021simdram, ferreira2022pluto, wu2022dram, deng2023dram, oliveira2024mimdram, gao2019computedram, gao2022fracdram, yuksel2024simultaneous, kubo2025pudtune, kubo2025mvdram, chang2016low, shivanandamurthy2021atria, afifi2024artemis, jahshan2024majork, mutlu2024memory, liu2025optipim, de2026count2multiply, garzon2026cadm, mutlu2025memory} is a new computing paradigm that leverages the analog properties of DRAM to enable massively parallel in-DRAM computation.
This work targets two representative PuD architectures. The first is \textit{SIMDRAM}~\cite{hajinazar2021simdram}, one of the most well-established PuD architectures, which supports \dtcr{3}{bulk bitwise}~\myNOT{} operations through the use of dual-contact cells \dtcr{3}{(originally introduced by Ambit~\cite{seshadri2017ambit})}. The second is \textit{Unmodified PuD}~\cite{kubo2025mvdram, kubo2025pudtune, jahshan2024majork, garzon2026cadm, gao2022fracdram, olgun2022pidram, yuksel2024functionally, yuksel2023pulsar, yuksel2024simultaneous}, which avoids modifications to the \dtcr{3}{DRAM chips} and is modeled after \dtcr{3}{computation capabilities that are demonstrated to be present in} commercial off-the-shelf (COTS) DRAM PIM~\cite{gao2019computedram, olgun2021quac, gao2022fracdram, yuksel2023pulsar, yuksel2024functionally, yuksel2024simultaneous, olgun2025dram, yuksel2025pudhammer, kubo2025pudtune, kubo2025mvdram, olgun2022pidram, mutlu2024memory, mutlu2025memory}. We note that \dtcr{3}{we do not aim to claim an execution model that can be immediately adopted in existing systems with COTS DRAM chips}. Instead, we choose this \dtcr{3}{Unmodified PuD architecture} as a candidate for future PuD systems that \dtcr{3}{can} offer the lowest manufacturing \dtcr{3}{cost and minimal changes to DRAM circuitry}. We experimentally confirm the practical feasibility of Unmodified PuD and obtain precise operation latencies through experiments on off-the-shelf DDR4 DRAM modules using DRAM Bender~\cite{olgun2023dram, safari2022drambender}, an FPGA-based custom memory control infrastructure~\cite{olgun2023dram,safari2022drambender, hassan2017softmc,safari2017softmc} (Figure~\ref{fig:fpga}).

PuD computation \dtcr{3}{consists of} repeated invocations of two primitive PuD operations: \RowCopy{} and \MAJ{3}.
Each PuD operation is realized by issuing a dedicated sequence of DRAM commands such as \ACT{} and \PRE{}~\cite{seshadri2017ambit, hajinazar2021simdram, gao2019computedram, yuksel2024simultaneous}.
\RowCopy{} transfers data between rows within the same subarray, enabling efficient bulk data movement inside DRAM \dtcr{3}{via two consecutive activations within the same subarray in quick succession}.
The \MAJ{3} operation activates multiple rows to compute a three-input \dtcr{3}{bulk bitwise} majority function. 
By using one of the constant rows (rows filled entirely with 0s or 1s) as a fixed input, \MAJ{3} realizes \myAND{} and \myOR{} operations, serving as a fundamental logic primitive for PuD.

Different PuD architectures support \MAJ{3} in different ways. SIMDRAM \dtcr{3}{(and Ambit)} enables the simultaneous activation of three rows within a designated row group \dtcr{3}{in each subarray}. 
In contrast, \dtcr{3}{Unmodified DRAM cannot simultaneously activate exactly three rows. Prior work~\cite{olgun2021quac, yuksel2023pulsar, yuksel2024simultaneous,olgun2025dram,kubo2025pudtune,yuksel2025pudhammer} observes that COTS DRAM chips support simultaneous activation of four rows but not three, and attributes this limitation to the hierarchical row decoder design. To perform \MAJ{3}, Unmodified DRAM instead activates four rows through a sequence of \ACT{} and \PRE{} commands.
Before the four-row activation, one of the four rows is set to an intermediate voltage level using the \Frac{} operation~\cite{gao2022fracdram, kubo2025pudtune}, effectively neutralizing its contribution to the majority vote and making the result equivalent to a three-input majority. Because Unmodified DRAM does not support logical \myNOT{} natively, prior work maintains both a value and its logical complement throughout the computation, achieving functional completeness without a dedicated \myNOT{} operation~\cite{gao2019computedram, kubo2025mvdram}.}

\begin{figure}[htbp]
    \centering
    \includegraphics[width=\columnwidth]{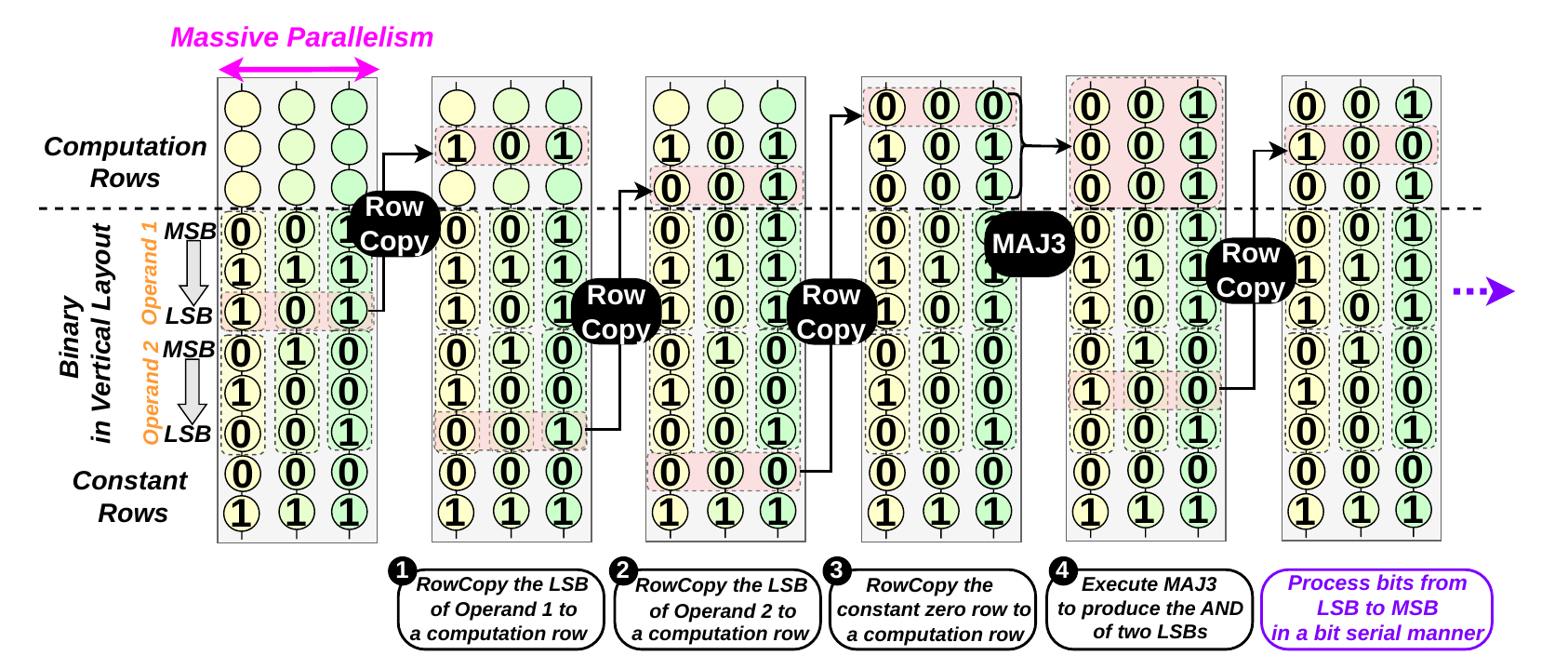}
    \caption{\dtcr{2}{Bit-serial-based PuD arithmetic.}}
    \label{fig:vertical}
\end{figure}

Prior PuD performs arithmetic by vertically aligning the binary representations of operands within a column and executing a sequence of PuD operations in a bit-serial manner~\cite{seshadri2017ambit, hajinazar2021simdram, oliveira2024mimdram, oliveira2025proteus, liu2025optipim}. \dtcr{3}{Figure~\ref{fig:vertical} illustrates the first step of a multiplication between two operands, operand 1 and operand 2, whose bits are stored vertically in the same column.
To compute the \myAND{} of their LSBs, the PuD first uses \RowCopy{} operations to copy the two LSB bits (\bcircled{1} and \bcircled{2}) along with a constant row filled with zeros (\bcircled{3}) into a set of designated rows called compute rows, where \MAJ{3} can be performed. Executing \MAJ{3} over these three bits produces the \myAND{} of the two LSB bits (\bcircled{4}). By processing each bit from the LSB to the MSB in this bit-serial manner, PuD performs full arithmetic operations such as multiplication. Because DRAM applies the same command sequence across all columns within a bank, PuD performs the same arithmetic operation concurrently on every column, enabling massive parallelism.}

\section{Motivation}\label{sec:mot}

\subsection{Bottleneck Profiling of Vector-Scalar Comparisons}\label{sec:mot:prof}
\begin{figure}[htbp]
    \centering
    \vspace{-1.5em}
    \includegraphics[width=0.6\columnwidth]{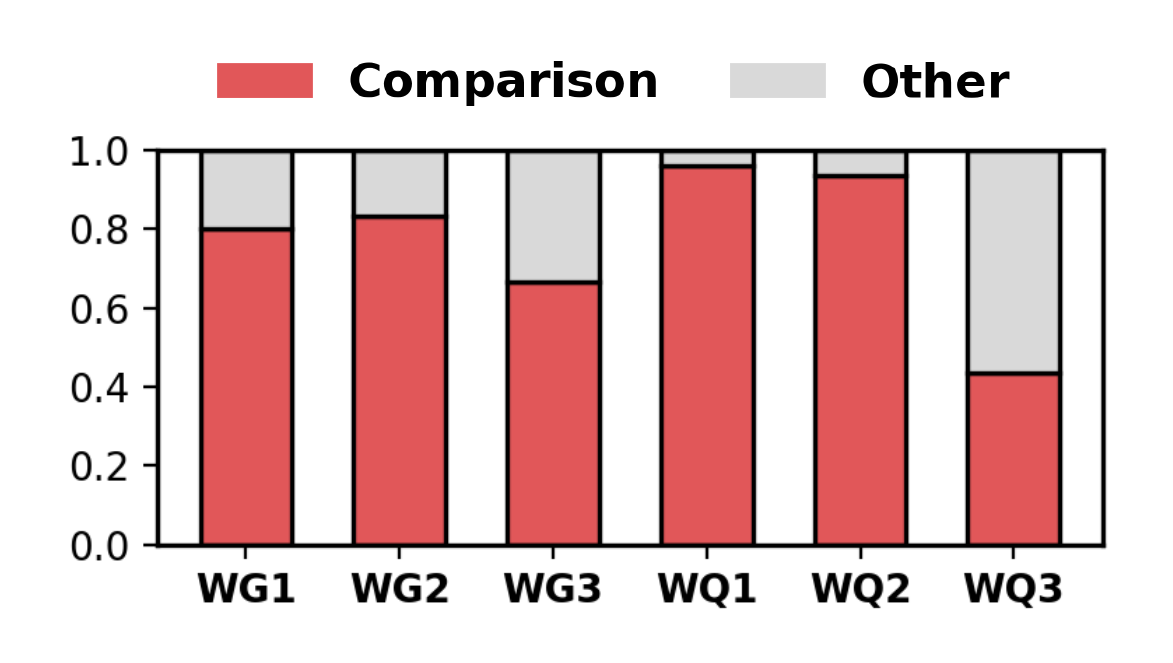}
    \vspace{-1.0em}
    \caption{Fraction of execution time spent on comparison.}
    \vspace{-0.5em}
    \label{fig:profile}
\end{figure}

Vector-scalar comparisons are widely used across a broad range of applications that support the foundation of modern society, from query processing in in-memory databases to inference in decision tree ensembles for machine learning. In many of these workloads, accelerating the comparison step is particularly effective, as it has a substantial impact on overall performance~\cite{khataei2025treelut,willhalm2009simd,li2013bitweaving}.
Our profiling results \dtcr{6}{on real CPU systems (e.g., Intel Core i7-9700K ~\cite{intel2018i79700k})} show that comparison operations account for up to 96\% of the total execution time in query processing and up to 81\% in GBDT inference\footnote{For GBDT inference, the comparison step includes producing leaf addresses based on comparisons between feature values and node thresholds.} (see~\S\ref{sec:app} for the system configuration and workload details). Figure~\ref{fig:profile} illustrates the \dtcr{3}{fraction of execution time spent on} comparison operations for a subset of workloads evaluated in~\S\ref{sec:app}. WG1--WG3 correspond to GBDT inference: WG1 uses depth 12 with a batch size 1024, WG2 uses depth 8 with a batch size 1024, and WG3 uses depth 10 with a batch size 64. WQ1--WQ3 correspond to predicate evaluation queries Q2--Q4 in \S\ref{sec:app:que}.

Due to the very low computational intensity of the comparison operation, its performance is bottlenecked by off-chip data movement for large datasets~\cite{li2013bitweaving}. We confirm this by profiling an optimized comparison kernel~\cite{li2013bitweaving}, which employs a transposed layout for vector–scalar comparison: loading the vector data from DRAM alone accounts for the vast majority of the total execution time. This shows that further improving comparison performance requires alleviating the memory-bandwidth bottleneck.

\subsection{Data Movement Reduction via PuD Execution}\label{sec:mot:ana}
\begin{figure*}[t]
    \centering
    \includegraphics[width=\textwidth]{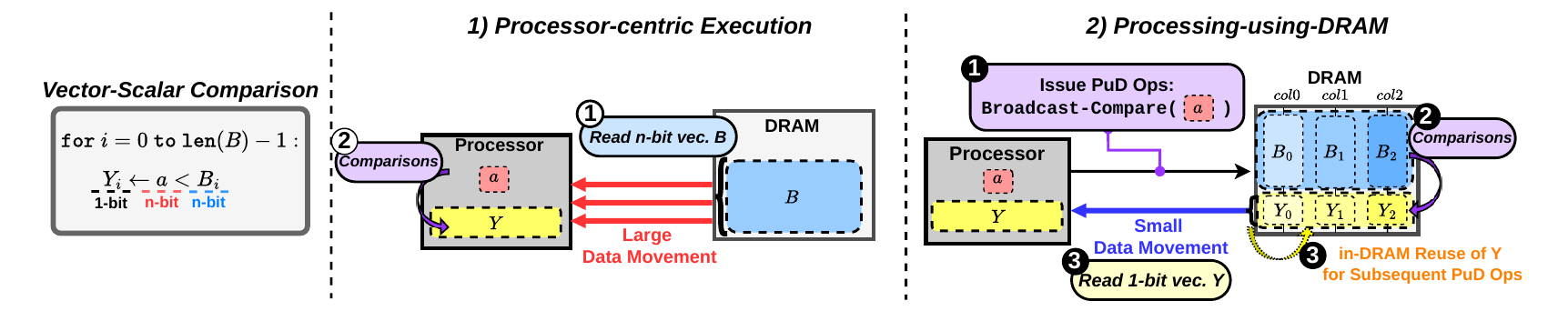}
    \caption{\dtcr{2}{PuD execution effectively reduces data movement for vector-scalar comparison.}}
    \label{fig:analysis}
\end{figure*}

Figure~\ref{fig:analysis} illustrates that PuD explicitly reduces off-chip data movement compared to the processor-centric approach, leveraging the arithmetic nature of vector–scalar comparisons. We consider a scenario where \dtcr{3}{a vector \B{} of $n$-bit elements} is stored in DRAM, and the host processor holds a scalar value $a$. The goal is to compute a comparison $a < B_i$ for each $n$-bit element $B_i$ and obtain the resulting 1-bit output vector \Y{}.

We begin by describing the execution on the processor. Assuming that the vector \B{} is much larger than the processor’s cache, the processor must first load \B{} from DRAM (\wcircled{1}). Then, the comparison is performed, and the processor obtains the output vector \Y{} (\wcircled{2}). Depending on the size of \B{} and the structure of the subsequent processing stages, \Y{} may also not fit in the cache. 
The lower bound on DRAM accesses is at least reading the $n$-bit vector \B{}.

In contrast, in PuD execution, the vector \B{} is stored in DRAM in a vertical layout. \dtcr{3}{To perform the comparison, each bit of the scalar $a$ must be initialized across all columns so that PuD operations can process it against the corresponding bits of \B{}. A naive approach would be to \WRITE{} the scalar value into every column, but this would incur the same off-chip data movement that PuD aims to eliminate. Instead, following prior approaches~\cite{kubo2025mvdram,de2026count2multiply}, PuD can prepare constant rows filled entirely with $0$s or $1$s in advance across all columns, and then initialize each bit of $a$ by selecting the appropriate constant row and copying it using \RowCopy{}~\cite{seshadri2013rowclone, seshadri2017ambit, gao2019computedram, olgun2022pidram, yuksel2024simultaneous}. For example, if $a = 3$ ($a_2 a_1 a_0 = 011$ in binary), it copies the all-$1$s row for $a_0$ and $a_1$ and the all-$0$s row for $a_2$. In this way, PuD dynamically issues PuD operations according to the value of $a$ (\bcircled{1}), and the comparison is executed entirely inside DRAM without any off-chip data movement of $a$ (\bcircled{2}). The output vector \Y{} with 1-bit elements is produced directly within DRAM, and the host processor reads \Y{} only when it is needed for subsequent processing (\bcircled{3}).}

Crucially, the output bitmap \Y{} is often combined with other comparison-result bitmaps (e.g., via \myAND{}/\myOR{} \dtcr{6}{operations}) in real workloads (see also~\S\ref{sec:app}). PuD can execute these bitwise operations directly \dtcr{6}{in DRAM} at high throughput, avoiding transfers of \Y{} to the processor until the final result is needed (described as the yellow arrow in Figure~\ref{fig:analysis}).

In summary, processor-side execution involves transferring at least the full $n$-bit vector \B{}, assuming that \B{} is much larger than the processor cache. In contrast, PuD execution only requires transferring the 1-bit vector \Y{} of the same length, or potentially no transfer of \Y{} at all if the subsequent bitwise reduction is handled by PuD. Therefore, for large vectors, PuD can reduce the data transfer volume by a factor of at least $n$. This advantage comes from three key properties of vector–scalar comparison: (i) the scalar value is shared across all elements, (ii) the output is a vector with 1-bit elements \dtcr{6}{(i.e., a bitmap)}, 
and (iii) the output bitmap is often combined with other bitmaps in downstream processing.

\subsection{Limitations of Bit-Serial Comparison in PuD}\label{sec:mot:lim}
\begin{figure}[htbp]
    \centering
    \includegraphics[width=\columnwidth]{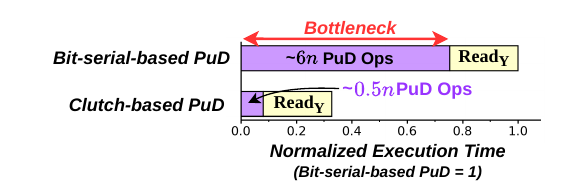}
    \caption{Execution time breakdown of vector–scalar comparison at 32-bit precision on PuD.}
    \label{fig:comp_comp}
\end{figure}

While PuD can explicitly reduce data transfer for vector–scalar comparisons, the bit-serial comparison~\cite{seshadri2017ambit, hajinazar2021simdram, oliveira2024mimdram} still limits the end-to-end application-level speedup over processor-centric execution.
This limitation arises mainly because, as illustrated in Figure~\ref{fig:comp_comp}, the bit-serial approach requires ${\sim}\,6n$ PuD operations for an unsigned $n$-bit comparison per DRAM bank in Unmodified PuD, and ${\sim}\,4n$ in SIMDRAM~\cite{kubo2025mvdram, seshadri2017ambit, hajinazar2021simdram}.~\footnote{The ${\sim}4n$ operations on SIMDRAM include \RowCopy{} operations to initialize the scalar value \smallA{} within the subarray from constant rows. The ${\sim}6n$ operations on Unmodified PuD additionally include \RowCopy{} to a neutral row and a \Frac{} operation per step.}
\dtcr{3}{This overhead makes it difficult to outperform the processor-based approach when parallelism is limited or when additional overhead from non-comparison operations arises in real-world applications (see Figures~\ref{fig:gbdt_config} and~\ref{fig:ts_q5_gpu}).}

Our analysis further reveals that the latency of comparisons is dominated by the number of PuD operations. 
\dtcr{3}{While PuD eliminates the need to move the vector \B{} from DRAM to the processor, the large number of PuD operations required for bit-serial comparison becomes the performance bottleneck.} 
\dtcr{3}{For example, as shown in Figure~\ref{fig:comp_comp}, PuD operations account for 76\% of the total latency in 32-bit comparisons on a system configuration summarized in Table~\ref{tab:spec:intel}, clearly indicating that PuD operations, not data movement, are the primary limiter.}
Therefore, reducing the number of PuD operations is a key challenge, and addressing it is critical to unlocking the full performance potential of PuD for comparison-intensive applications.



\section{Clutch}
To reduce the large number of PuD operations incurred by bit-serial execution, we present Clutch, a novel comparison algorithm and data representation for PuD.

\subsection{Lookup Table-based Approach}
\begin{figure}[htbp]
    \centering
    \includegraphics[width=\columnwidth]{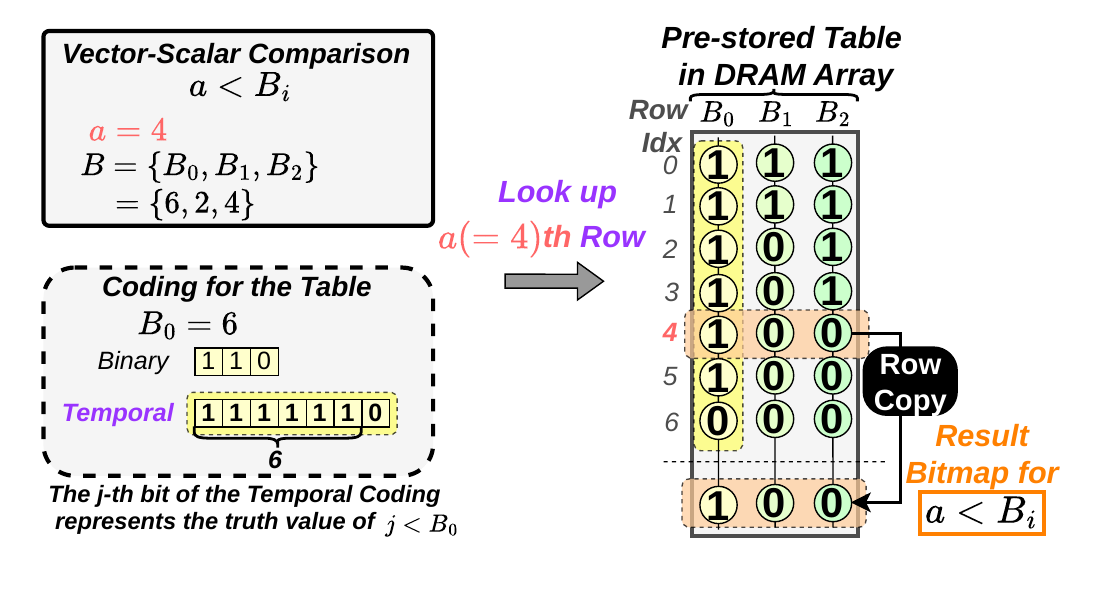}
    \vspace{-1.0em}
    \caption{\dtcr{2}{Lookup table-based comparison via temporal coding.}}
    \vspace{-0.5em}
    \label{fig:clutch-naive}
\end{figure}

Instead of representing each element of \B{} in binary, Clutch uses \emph{temporal coding}~\cite{madhavan2014race,wu2020ugemm}, where a value $v$ is represented as a bitstream of $v$ leading ones followed by zeros. \dtcr{3}{A key property of this encoding is that the $i$-th bit equals the truth value of $i < v$. As illustrated in Figure~\ref{fig:clutch-naive}, when vector elements are stored in this format across the columns of a DRAM subarray, each column stores one vector element, and row $a$ directly contains the output bitmap of the vector–scalar comparison $a < B_i$ for all elements $B_i$.} 

\dtcr{3}{The data encoded with temporal coding is either converted offline and loaded into DRAM, or converted from binary by the host processor and stored in DRAM prior to execution (we discuss the cost of this conversion in \S\ref{sec:app}). At runtime, the host processor dynamically issues PuD operations based on the scalar value \smallA{}, executing the vector–scalar comparison within the DRAM subarray (we discuss the system design for dynamically issuing PuD operations in \S\ref{sec:dis:sys}).}

While this lookup table-based approach offers extremely high throughput, it requires a large number of rows ($2^n - 1$ rows) for $n$-bit values, which poses practical challenges when mapped to DRAM rows. Since typical PuD architectures have DRAM subarrays with \dtcr{3}{e.g.,} 1024 rows~\cite{seshadri2017ambit, hajinazar2021simdram, oliveira2024mimdram, oliveira2025proteus}, supporting $n = 16$ or $n = 32$ far exceeds the available row budget. Moreover, as described in~\S\ref{sec:app:que}, some applications benefit from placing multiple vectors in the same subarray columns to enable in-DRAM reductions over multiple result bitmaps. In such cases, even for $n \leq 8$, a more compact design is required. Therefore, we need a new approach that preserves the high throughput of lookup-based comparisons while significantly reducing the required number of rows.

\subsection{Divide-and-Conquer Approach for Comparisons}
\begin{figure}[htbp]
    \centering
    \vspace{-1.0em}
    \includegraphics[width=\columnwidth]{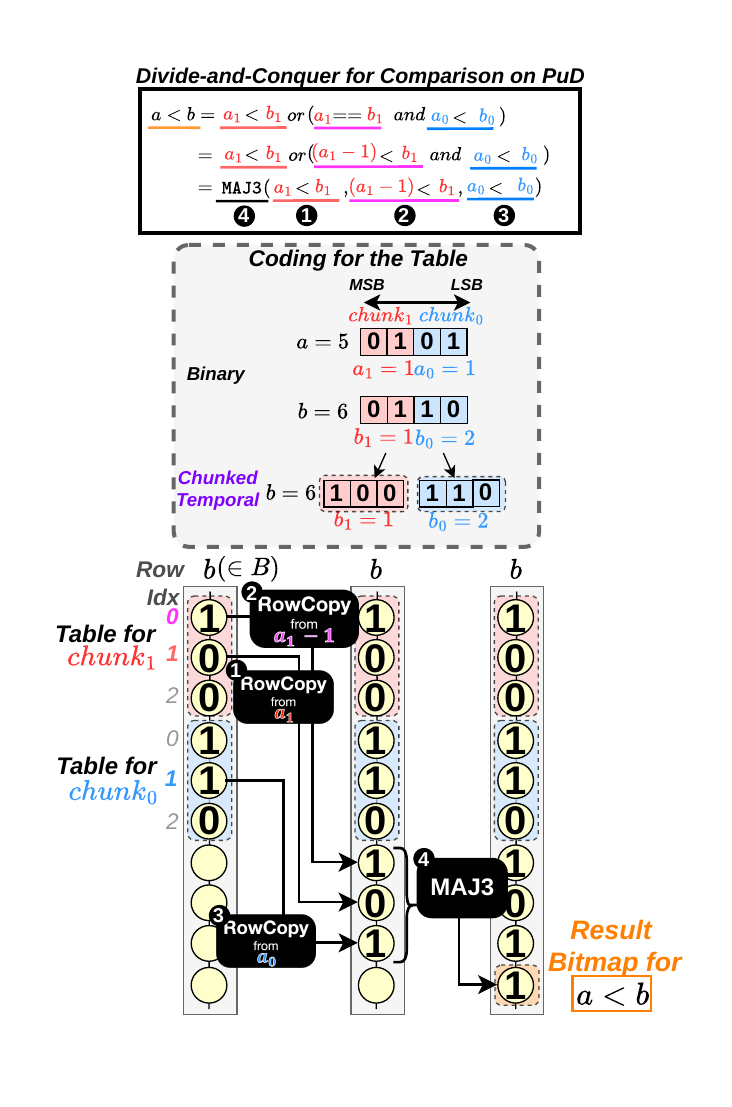}
    \vspace{-4.0em}
    \caption{Clutch encoding and algorithm example.}
    \vspace{-0.0em}
    \label{fig:clutch-chunk}
\end{figure}

To support scalable and memory-efficient lookup table-based comparisons, Clutch adopts a divide-and-conquer approach. Our key idea is that a full-width comparison can be decomposed into smaller sub-comparisons over bit chunks. Specifically, Clutch partitions each binary operand into multiple multi-bit chunks, compares each chunk independently using a compact lookup table encoded with temporal coding, and then merges the partial results through a procedure that propagates carry information across chunks, optimized for PuD execution.
Figure~\ref{fig:clutch-chunk} illustrates Clutch's ~\dtcr{6}{divide-and-conquer approach} for a 4-bit binary number split into two chunks: an MSB chunk and an LSB chunk. For each chunk, a straightforward lookup-table comparison under temporal coding produces a per-chunk result.
To merge the per-chunk results efficiently on PuD, we rewrite the comparison expression $a < b$.
Let $a_1$ and $b_1$ denote the MSB chunks and $a_0$ and $b_0$ denote the LSB chunks.
A direct reformulation yields:
$$
a < b \quad\Leftrightarrow\quad a_1 < b_1 \;\text{or}\; \bigl((a_1 == b_1) \;\text{and}\; a_0 < b_0\bigr)
$$
Because the equality term $(a_1 == b_1)$ is not directly available from the chunk-level comparisons, we substitute it as follows:
$$
a < b \quad\Leftrightarrow\quad a_1 < b_1 \;\text{or}\; \bigl((a_1 - 1) < b_1 \;\text{and}\; a_0 < b_0\bigr)
$$
Two steps justify this transformation.
First, relaxing $a_1 == b_1$ to $a_1 \le b_1$ only introduces the case $a_1 < b_1$, which is already covered by the first term, so the overall expression remains equivalent.
Second, for integer comparisons, $a_1 \le b_1$ can be rewritten as $(a_1 - 1) < b_1$.
With this form, the comparison $(a_1 - 1) < b_1$ can be evaluated using the same lookup table as $a_1 < b_1$, simply by using $(a_1 - 1)$ as the index instead of $a_1$.
\dtcr{7}{As shown in Figure~\ref{fig:clutch-chunk}, Clutch evaluates three sub-comparisons: $a_1 < b_1$ (\bcircled{1}), $(a_1 - 1) < b_1$ (\bcircled{2}), and $a_0 < b_0$ (\bcircled{3}), each via a single lookup-table access, and then combines them into the final result using a single \MAJ{3} operation (\bcircled{4}).}
\dtcr{7}{The Boolean expression can be simplified to a single \MAJ{3}  because} if $a_1 < b_1$ holds then $(a_1 - 1) < b_1$ also holds.

Although Figure~\ref{fig:clutch-chunk} shows the two-chunk case, this divide-and-conquer algorithm can be applied recursively from the LSB side when the number of chunks exceeds two, producing a PuD-optimized comparison algorithm.
Algorithm~\ref{alg:conquer} shows the general algorithm of Clutch to compute vector–scalar comparisons. The algorithm processes chunks from the LSB chunk to the MSB chunk (lines 2--13), progressively merging per-chunk comparison results using a single \MAJ{3} operation per chunk (line 12).
The array $cp$ stores the starting row index of the lookup table for each chunk, so that $\mathrm{row}[a_j + cp[j]]$ retrieves the result of comparing chunk value $a_j$ against the corresponding chunk of each vector element $B_i$. The host processor retains $cp$ and the chunk values $a_0, \dots, a_{C-1}$ ($C$ denotes the number of chunks), and dynamically issues PuD operations based on these values.


\begin{algorithm}[h]
\small
\caption{PuD-Optimized Clutch Algorithm}\label{alg:conquer}
\begin{algorithmic}[1]
  \REQUIRE $C$: the number of chunks
  \REQUIRE $a_0,\dots,a_{C-1}$: individual chunks of scalar $a$ (from LSB to MSB)
  \REQUIRE $k_0,\dots,k_{C-1}$: bit-width of each chunk
  \REQUIRE $cp[0\dots C-1]$: starting row index of the table for each chunk
  \ENSURE comparison result $L$ ($L = 1$ if $a < B_i$ for each column $i$)
  \IF{$a_0 = 2^{k_0} - 1$}
    \STATE $L \gets 0$ // $a_0 < b_0$ is always false when $a_0 = 2^{k_0} - 1$
  \ELSE
    \STATE $L \gets \mathrm{row}[\,a_0 + cp[0]\,]$ // $L \gets (a_0 < b_0)$
  \ENDIF
  \FOR{$j = 1$ \textbf{to} $C-1$}
    \IF{$a_j = 2^{k_j} - 1$}
      \STATE $lt \gets 0$ // $a_j < b_j$ is always false when $a_j = 2^{k_j} - 1$
    \ELSE
      \STATE $lt \gets \mathrm{row}[\,a_j + cp[j]\,]$ // $lt \gets (a_j < b_j)$
    \ENDIF
    \IF{$a_j = 0$}
      \STATE $le \gets 1$ // $a_j \le b_j$ is always true when $a_j = 0$
    \ELSE
      \STATE $le \gets \mathrm{row}[\,a_j - 1 + cp[j]\,]$ // $le \gets (a_j - 1 < b_j) = (a_j \le b_j)$
    \ENDIF
    \STATE $L \gets \mathrm{MAJ3}(L,\,lt,\,le)$ // Combine chunks: $lt \;\text{or}\; (le \;\text{and}\; L)$
  \ENDFOR
  \RETURN $L$
\end{algorithmic}
\end{algorithm}

The algorithm first initializes $L$ with the LSB chunk comparison result (lines 1--5). When $a_0 = 2^{k_0} - 1$, the lookup table access would exceed the table boundary, but since $a_0 < b_0$ is always false for the maximum chunk value, the algorithm sets $L = 0$ using a constant-zero row (line 2). For each subsequent chunk $j \ge 1$ (lines 6--18), the algorithm computes two values: $lt = (a_j < b_j)$ (lines 7--11) and $le = (a_j \le b_j)$ (lines 12--16). The value $lt$ is obtained by looking up $a_j$ in the lookup table (line 10), with the same maximum-value boundary case handled by setting $lt = 0$ (line 8). The value $le$ is obtained by looking up $(a_j - 1)$ in the same table (line 15). When $a_j = 0$, this access would underflow, but since $a_j \le b_j$ is always true, the algorithm sets $le = 1$ using a constant-one row (line 13).
Finally, the algorithm combines $L$, $lt$, and $le$ via $\mathrm{MAJ3}$ to propagate the comparison result from lower chunks to the current chunk (line 17). This operation computes $lt \;\text{or}\; (le \;\text{and}\; L)$, matching the reformulated comparison expression described above. Notably, Clutch does \emph{not} require a logical \myNOT{} operation \dtcr{7}{for} any step. As a result, even on Unmodified DRAM, Clutch does \emph{not} need to maintain the logical complement of a value.

\begin{figure}[htbp]
    \centering
    \includegraphics[width=0.85\columnwidth]{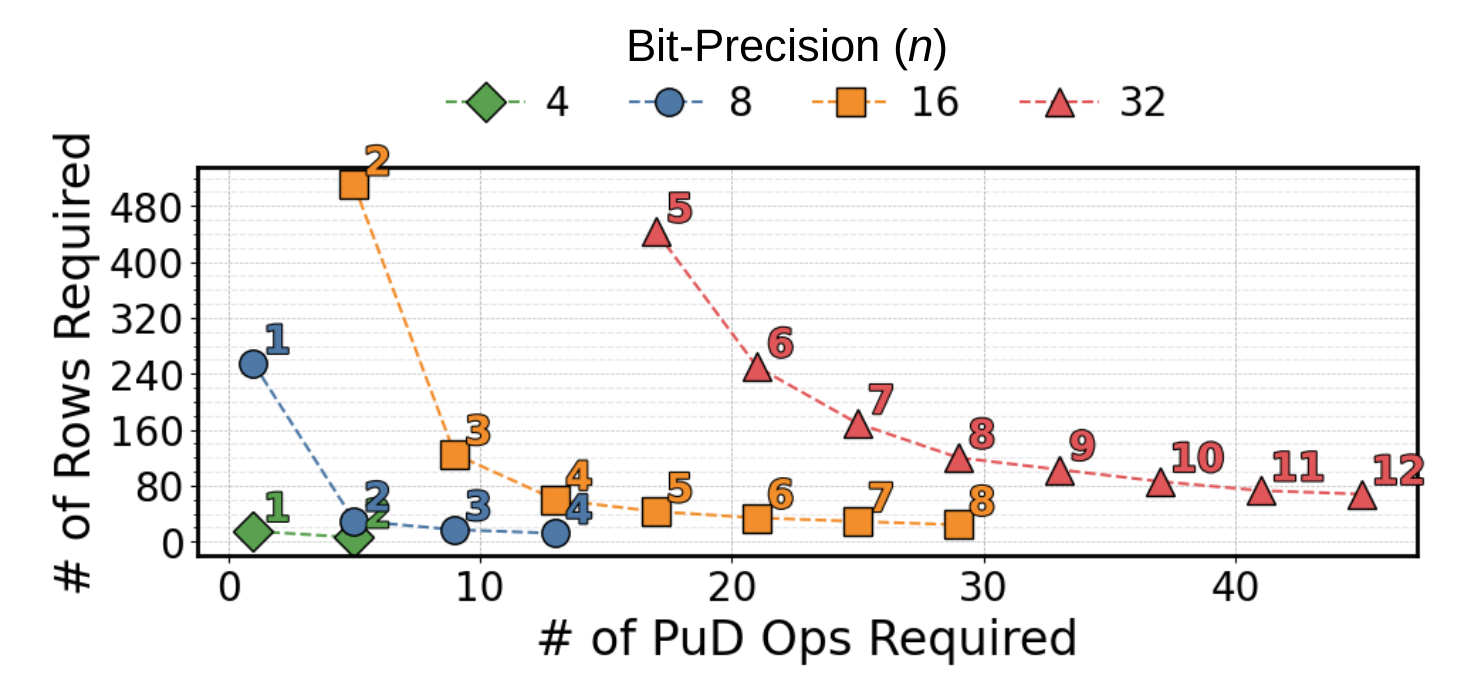}
    \caption{Tradeoff between DRAM row usage and the number of PuD operations in Clutch. \dtcr{6}{Each point represents a different chunk count on Unmodified DRAM, annotated with the number of chunks used.}}
    \vspace{-0.5em}
    \label{fig:trade_apa}
\end{figure}

The number of PuD operations in Clutch depends only on the number of chunks $C$, not the operand bit-precision $n$. Clutch enables a flexible tradeoff between throughput and memory footprint by adjusting the number of chunks. 
Since each chunk's lookup table has $2^k - 1$ rows for a $k$-bit chunk and this grows exponentially with $k$, the total number of rows is minimized when the $n$ bits are distributed as evenly as possible across the $C$ chunks. 
Figure~\ref{fig:trade_apa} illustrates this tradeoff space across different operand bit-precisions ($n =$ 4, 8, 16, 32) and chunk counts for Unmodified DRAM, where each point represents a different chunk count annotated with the number of chunks used. 
For example, for 32-bit values with five chunks (chunk sizes of 6, 6, 6, 7, 7 bits, requiring $63+63+63+127+127 = 443$ rows), Clutch executes vector–scalar comparison in only 17 PuD operations on Unmodified DRAM. This yields over an order-of-magnitude reduction in PuD operations compared to the state-of-the-art bit-serial approach~\cite{seshadri2017ambit,hajinazar2021simdram,oliveira2024mimdram}.

\section{Evaluation}\label{sec:eva}
In~\S\ref{sec:eva} and~\S\ref{sec:app}, we compare processor-centric execution and PuD-based execution under the same processor and DRAM configuration to ensure a fair comparison. In~\S\ref{sec:eva}, we focus on the end-to-end performance of a vector–scalar comparison kernel, where a scalar value resides on the processor and the vector is resident in DRAM. 
We quantitatively evaluate the performance of two comparison methods, \dtcr{3}{the state-of-the-art} bit-serial execution (SIMDRAM and Ambit)~\cite{hajinazar2021simdram,seshadri2017ambit} and Clutch. Table~\ref{tab:spec:intel} summarizes the processor and DRAM system configurations used in~\S\ref{sec:eva}.\footnote{The workloads evaluated in \S\ref{sec:eva} and \S\ref{sec:app:que} are memory-bandwidth bound~\cite{li2013bitweaving}, so the processor generation has little impact on the throughput. We use DDR4-2666~\cite{jedec2012ddr4} because it enables \dtcr{3}{us to obtain} hardware-verified PuD operation latencies using our FPGA platform (see Figure~\ref{fig:fpga}).} 
These configurations are shared across all evaluated execution methods and algorithms.

We evaluate two representative PuD architectures, which we also use in~\S\ref{sec:app}.

\noindent\textit{\textbf{1) Unmodified PuD}}:
A COTS DRAM based PuD architecture that does \emph{not} require modifications to the DRAM circuitry~\cite{kubo2025mvdram, kubo2025pudtune, jahshan2024majork, garzon2026cadm, gao2022fracdram, olgun2022pidram, yuksel2024functionally, yuksel2023pulsar, yuksel2024simultaneous}. This architecture is modeled after COTS-DRAM chips~\cite{gao2019computedram,olgun2021quac,gao2022fracdram,yuksel2023pulsar,yuksel2024functionally,yuksel2024simultaneous,olgun2025dram,yuksel2025pudhammer,kubo2025pudtune,kubo2025mvdram} and realizes the \MAJ3{} operation using the \Frac{} operation~\cite{gao2022fracdram} and fixed four-row activation~\cite{yuksel2024simultaneous,kubo2025mvdram}.
We note that \dtcr{3}{we do not aim to claim an execution model that can be immediately adopted in existing systems with COTS DRAM chips}. 
Instead, we choose this Unmodified PuD architecture as a candidate for future PuD systems \dtcr{6}{that} can offer the lowest manufacturing cost and minimal changes to DRAM circuitry.
We validate its practicality and obtain precise, hardware-verified latencies of PuD operations on off-the-shelf DDR4 DRAM modules using DRAM Bender~\cite{olgun2023dram,safari2022drambender} (which is based on SoftMC~\cite{hassan2017softmc,safari2017softmc}), an FPGA-based custom memory control infrastructure (see Figure~\ref{fig:fpga}).

\noindent\textit{\textbf{2) Modified PuD}} (SIMDRAM):
A PuD architecture that adopts the same subarray structure as SIMDRAM~\cite{hajinazar2021simdram}, also used in several subsequent works~\cite{liu2025optipim, de2026count2multiply}. The DRAM cell array is equipped with dual-contact cells (originally introduced by Ambit~\cite{seshadri2017ambit}) to support in-DRAM bulk bitwise \myNOT{} operations. This Modified PuD architecture implements the \MAJ3{} operation via triple-row activations among a set of reserved rows \dtcr{6}{(i.e., compute rows)} within each subarray.

Following prior work~\cite{hajinazar2021simdram, seshadri2017ambit, liu2025optipim}, we assume each DRAM module contains 16 banks, and PuD is enabled in one subarray per bank, with each subarray containing 1024 rows. Under the configuration shown in Table~\ref{tab:spec:intel}, this yields a column-level parallelism of 64K columns $\times$ 16 banks $\times$ 2 \dtcr{3}{DIMMs per channel} $\times$ 2 channels. For precise evaluation, we do \emph{not} simply scale single-bank performance by a factor of 16. 
\dtcr{6}{Instead, we derive a cycle-accurate latency from DRAM command sequences that explicitly model DRAM bank-level parallelism (BLP). 
Note that our evaluation does \emph{not} exploit subarray-level parallelism (SALP)~\cite{kim2012case}. If SALP is further exploited for PuD computation, PuD parallelism would be much higher, as demonstrated in prior work~\cite{mutlu2024memory, oliveira2025proteus}.}

Our evaluation follows prevailing methodology in PuD evaluation~\cite{oliveira2022accelerating, liu2025optipim, hajinazar2021simdram, kubo2025mvdram, oliveira2024mimdram, oliveira2025proteus}. We first run applications on real machines and profile their execution at kernel granularity. For kernels that are accelerated by PuD, we subtract the measured CPU execution time of the kernel and replace it with the analytically derived PuD-based computation time inside DRAM. We then add the execution time of all other processing, without overlapping it with the PuD computation time. Since Clutch’s comparison results remain in DRAM rather than being cached in the CPU, we explicitly add the time to transfer the output bitmaps from DRAM back to the processor, yielding a conservative \dtcr{3}{(i.e., favoring baseline processor-centric systems)} end-to-end performance estimate.

The execution time on the processor is measured on real hardware \dtcr{3}{(see Table~\ref{tab:spec:intel} for the evaluation in ~\S\ref{sec:eva:per})}, and CPU power consumption is obtained using Intel RAPL~\cite{khan2018rapl}. For PuD-based computation inside DRAM, we analytically derive the execution time based on the sequence of DRAM commands required. The power consumption of PuD is estimated using data from CACTI 6.5~\cite{muralimanohar2009cacti}. For the energy cost of the PuD operations, we follow prior work~\cite{yuksel2024simultaneous, kubo2025mvdram} and assume that each additional simultaneously activated row increases the activation energy by 22\% relative to the single-row case, \dtcr{6}{as reported using real DRAM chips in ~\cite{yuksel2024simultaneous}}. For a fair comparison, we also account for the energy consumption of the host-side \dtcr{3}{processor execution} during PuD execution, assuming single-threaded processor power consumption.

\begin{table}[htbp]
\centering
\caption{Evaluated system configurations.}\label{tab:spec:intel}
\vspace{-1.0em}
\small
\setlength{\tabcolsep}{3pt}
\begin{tabular}{@{}ll@{}}
\toprule
\textbf{Main Memory} & 64 GB DDR4-2666 (4 × 16 GB)~\cite{jedec2012ddr4}, dual-channel; \\
\textbf{(COTS DRAM)} & Peak bandwidth: 42.6 GB/s; \\
& 2 ranks per DIMM, 16 banks per rank; \\
\midrule
\textbf{Processor} & Intel Core i7-9700K~\cite{intel2018i79700k}, 8 cores, up to 4.9 GHz; \\
\textbf{(Real CPU)} & 32 kB L1, 256 kB L2 per core; 12 MB shared L3; \\
\bottomrule
\end{tabular}
\end{table}

\subsection{Performance of Vector-Scalar Comparison}\label{sec:eva:per}
\begin{figure}[htbp]
    \centering
    \vspace{-1.0em}
    \includegraphics[width=0.9\columnwidth]{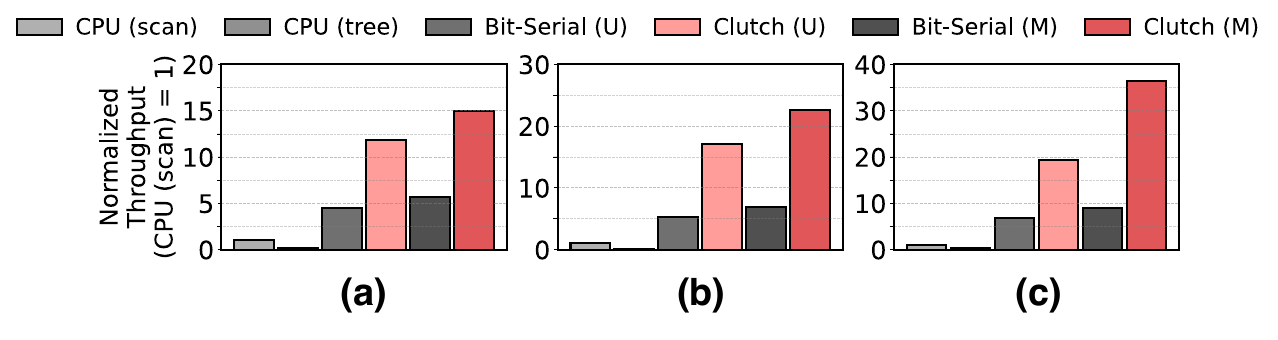}
    \vspace{-2.0em}
    \caption{Throughput of vector-scalar comparisons at (a)
8-bit precision, (b) 16-bit precision, and (c) 32-bit precision.}
    \vspace{-1.0em}
    \label{fig:comp_thr}
\end{figure}

\noindent Figure~\ref{fig:comp_thr} shows the performance comparison for vector–scalar comparisons with 256M elements. We evaluate six implementations: 1) \textit{CPU (scan)}, 2) \textit{CPU (tree)}, 3) \textit{Bit-Serial (U)} (the state-of-the-art bit-serial approach~\cite{hajinazar2021simdram, seshadri2017ambit} on Unmodified PuD), 4) \textit{Clutch (U)} (Clutch on Unmodified PuD), 5) \textit{Bit-Serial (M)} (the bit-serial approach on Modified PuD), and 6) \textit{Clutch (M)} (Clutch on Modified PuD).

For the main CPU baseline, labeled \emph{CPU (scan)} in the figure, we use an optimized kernel based on BitWeaving-V~\cite{li2013bitweaving}, a state-of-the-art technique designed to accelerate predicate evaluation in column scans for databases. BitWeaving-V addresses inefficiencies in conventional data layouts by storing data in a transposed, bit-sliced layout, where bits at the same position across many values are packed together, enabling parallel bitwise evaluation using SIMD instructions. In our implementation, BitWeaving-V outperforms other conventional scan layouts for all tested bit-precisions (8-bit, 16-bit, and 32-bit), and we therefore use \emph{CPU (scan)} as our primary processor-centric baseline.
In addition, we evaluate a tree-based CPU implementation, labeled \emph{CPU (tree)}, which reduces the number of comparison operations by organizing predicate thresholds in a search tree rather than scanning all predicate ranges sequentially.

Our evaluation explicitly includes the time required to transfer the final comparison result row back to the host \dtcr{3}{CPU for} PuD. For each bit-precision, Clutch uses the minimum number of chunks required to store a single value entirely within a single subarray (i.e., one chunk for 8-bit, two \dtcr{3}{chunks} for 16-bit, and five \dtcr{3}{chunks} for 32-bit), as shown in Figure~\ref{fig:trade_apa}.

Figure~\ref{fig:comp_thr} presents the throughput \dtcr{3}{of vector-scalar comparison on six systems, for three bit-precisions (8-bit, 16-bit, 32-bit)}.  Across all bit-precisions, CPU (scan) is consistently faster than CPU (tree) because the tree-based approach incurs irregular memory accesses when traversing the index structure and still requires linear-time memory accesses to build the bitmap. Consequently, we use CPU (scan) as the CPU reference for the speedup numbers reported. 

\begin{figure}[H]
    \centering
    \includegraphics[width=0.9\columnwidth]{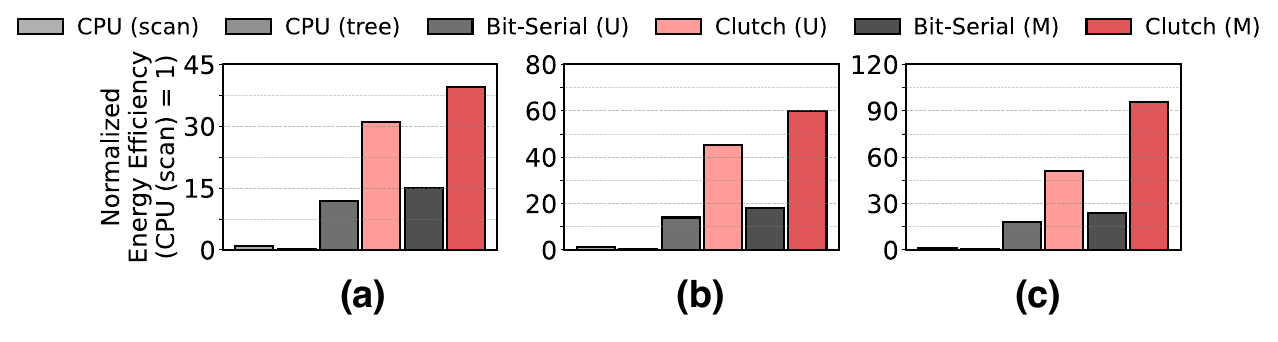}
    \vspace{-2.0em}
    \caption{Energy efficiency of vector-scalar comparisons at (a)
8-bit precision, (b) 16-bit precision, and (c) 32-bit precision.}
    \vspace{-1.0em}
    \label{fig:comp_ene}
\end{figure}

Clutch provides higher throughput as bit-precision increases, realizing up to 36$\times$ \dtcr{3}{(20$\times$ on average)} speedup over the CPU and up to 4.1$\times$ \dtcr{3}{(3.1$\times$ on average)} over bit-serial PuD. This trend is explained in Figure~\ref{fig:analysis}: CPU execution reads full-width operands from memory and is constrained by memory bandwidth, whereas PuD reads back only a result 1-bit-per-element bitmap. Consequently, PuD's relative data-movement savings grow with the operand bit-precision. In addition, Clutch substantially reduces the number of PuD operations compared to bit-serial PuD, directly alleviating the dominant performance bottleneck of bit-serial PuD (see Figure~\ref{fig:comp_comp}).
Figure~\ref{fig:comp_ene} presents the energy efficiency, \dtcr{3}{defined as the number of comparisons completed per unit of energy consumption,} relative to CPU (scan). Clutch improves energy efficiency by up to 96$\times$ \dtcr{3}{(54$\times$ on average)} compared to the CPU and up to 4.2$\times$ \dtcr{3}{(3.1$\times$ on average)} compared to bit-serial PuD.

\section{Accelerating Applications}\label{sec:app}
We demonstrate the applicability of Clutch to real-world workloads by evaluating \dtcr{6}{it} on two representative cases that benefit from the acceleration of comparisons: inference on \textbf{Gradient Boosting Decision Tree (GBDT)}, \dtcr{6}{which is widely used in finance, healthcare, and web services,} and \textbf{predicate evaluation}, which \dtcr{3}{is} commonly used in query processing \dtcr{3}{for} databases, scientific computing, and image processing.

\subsection{Gradient Boosting Decision Tree Inference}\label{sec:app:gbdt}
GBDT models~\cite{ke2017lightgbm, chen2016xgboost, prokhorenkova2018catboost, hancock2020catboost} are powerful ensemble models composed of multiple decision trees, demonstrating excellent performance in both classification and regression tasks on tabular data. GBDT has become a standard method in a wide range of industrial applications that involve tabular data~\cite{shwartz2022tabular, bojer2021kaggle}.
Compared to deep neural networks (DNNs), many comparative studies have shown that GBDT offers advantages in terms of accuracy, computational cost, and interpretability~\cite{shwartz2022tabular, shmuel2024comprehensive}. A recent large-scale benchmark study~\cite{shmuel2024comprehensive} that compared 14 tree-based and deep learning models on 111 tabular datasets reported that GBDT-based models occupied the top three positions.

Due to its lower computational cost \dtcr{3}{in} the inference phase compared to DNNs, GBDT is particularly widely used in resource-constrained edge devices~\cite{khataei2025treelut, tokuda2025df,zhang2021satellite}. Practical applications include remote sensing, anomaly detection, and predictive analytics on IoT sensor devices~\cite{alcolea2020inference, hu2018automated, zhang2021predictive, zhang2021using,zhang2021satellite, sun2020gradient}.
As a result, GBDT has become an essential component in edge AI systems, and further improvements in inference efficiency are actively pursued.

Among various implementations of GBDT, CatBoost~\cite{prokhorenkova2018catboost, hancock2020catboost} is known as one of the most powerful models. In the aforementioned benchmark study~\cite{shmuel2024comprehensive}, CatBoost ranked first on 19 individual datasets and achieved the best average rank of 4.9. CatBoost adopts a regular branching structure known as oblivious trees~\cite{prokhorenkova2018catboost, hancock2020catboost}, where all nodes at the same depth share the same feature index and threshold, as illustrated in Figure~\ref{fig:gbdt}. 
This property enables parallel execution of all $num\_trees \times num\_depth$ comparisons regardless of previous branching results. The final prediction is obtained by summing the leaf values reached in all trees.


\subsubsection{Implementation with Clutch}
Our contributions include a novel method for accelerating CatBoost inference under the PuD architecture. To our knowledge, this is the first demonstration of accelerating GBDT inference on PuM. Our mapping of CatBoost to PuD execution is based on our key insight that tree traversal can be reformulated as a sequence of vector–scalar comparisons followed by mask operations. Furthermore, in this mapping, the per-node bitmaps directly correspond to leaf addresses, minimizing the cost of reading out inference results.

\begin{figure}[htbp]
    \centering
    \includegraphics[width=\columnwidth]{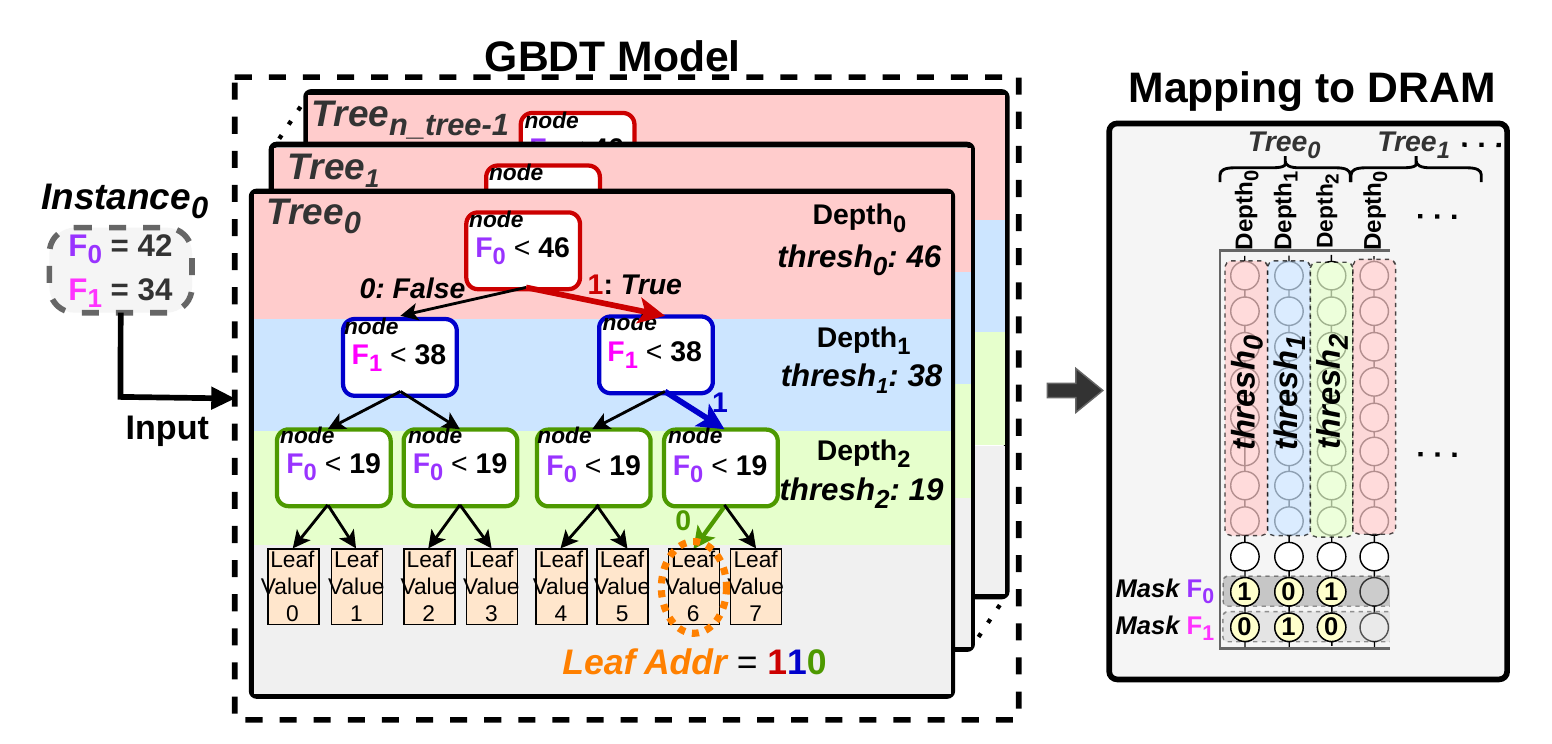}
    \caption{Mapping of GBDT trees to DRAM. }
    \label{fig:gbdt}
\end{figure}

\noindent\textit{\textbf{Layout.}} We map each decision-tree node to a single DRAM column, as illustrated in Figure~\ref{fig:gbdt}. Each column stores (i) the node’s threshold and (ii) a one-hot feature mask indicating which feature the node uses. Nodes are grouped by tree, and within each tree, columns are arranged in order of increasing depth so that nodes at the next depth are logically adjacent. 

\begin{figure}[htbp]
    \centering
    \vspace{-0.5em}
    \includegraphics[width=\columnwidth]{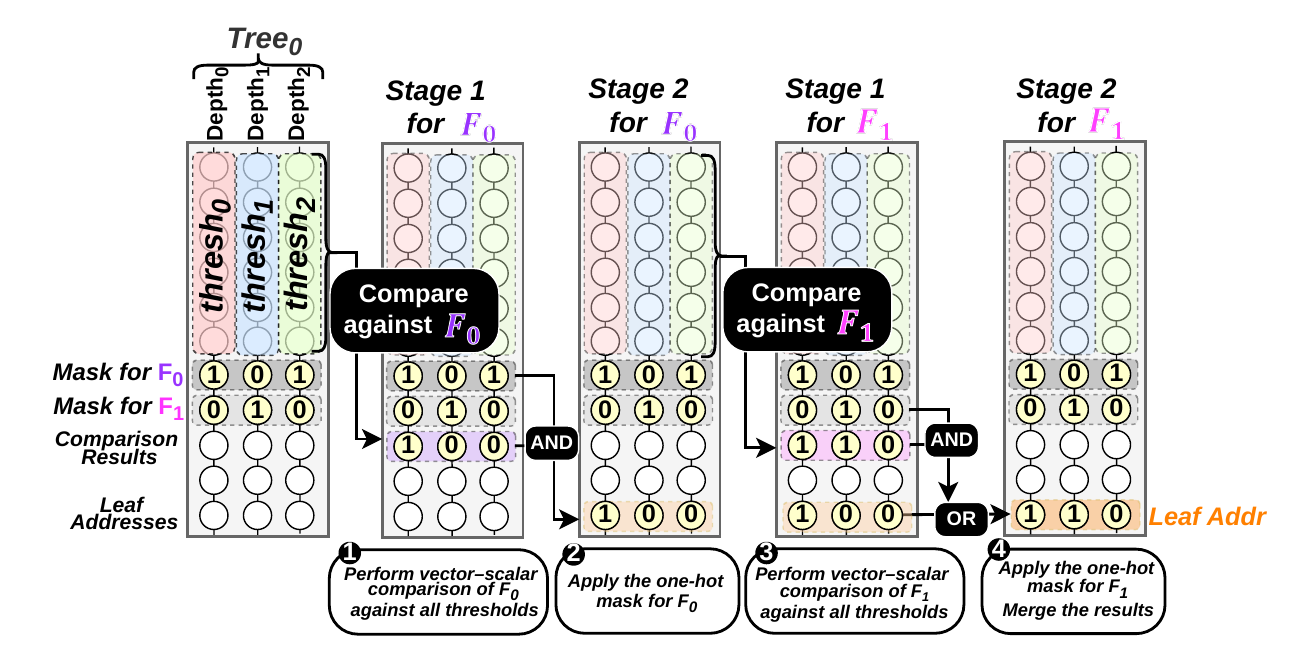}
    \vspace{-1.5em}
    \caption{Proposed GBDT inference flow on PuD.}
    \label{fig:palc_gbdt}
\end{figure}

\noindent\textit{\textbf{Execution flow.}} Figure~\ref{fig:palc_gbdt} shows the proposed execution flow for CatBoost inference on PuD. The algorithm processes one feature at a time \dtcr{6}{using a two-stage process}.
In \dtcr{6}{the first stage of processing a feature}, the host processor issues PuD operations based on the current feature value to perform a vector–scalar comparison against all node thresholds across all columns. This comparison is applied uniformly to all columns, regardless of which feature each node actually uses, yielding a 1-bit result for "(feature value) $<$ (node threshold)" in each column.
\dtcr{3}{In the second stage of processing a feature, the one-hot mask for the current feature is applied via a bitwise \myAND, so that only columns whose nodes use the current feature retain their comparison result; all others are cleared to zero. The masked result is then merged into the accumulated leaf address bitmap using a bitwise \myOR.}
\dtcr{6}{The algorithm repeats this two-stage process for every feature used by the model. Each DRAM bank processes one input instance, and multiple banks operate concurrently.}

\dtcr{7}{Figure~\ref{fig:palc_gbdt} illustrates this flow for a model with two features, $F_0$ and $F_1$, showing the four steps that result from applying the two-stage process to each feature.} \dtcr{6}{First, a vector–scalar comparison of $F_0$ is performed against all node thresholds (\bcircled{1}). The one-hot mask for $F_0$ is then applied via bitwise \myAND, retaining comparison results only for nodes that use $F_0$ (\bcircled{2}). Next, a vector–scalar comparison of $F_1$ is performed against all node thresholds (\bcircled{3}). Finally, the one-hot mask for $F_1$ is applied via bitwise \myAND, and the result is merged with the $F_0$ bitmap produced in \bcircled{2} via bitwise \myOR{} to update the leaf address bitmap (\bcircled{4}).}

\noindent\textit{\textbf{Leaf addresses in DRAM.}} After sweeping all features, each column holds the final comparison result for its node. Because nodes are laid out by depth, the comparison results across depths naturally form a binary encoding of the leaf address. For example, in Figure~\ref{fig:gbdt} and Figure~\ref{fig:palc_gbdt}, a tree of depth 3 has leaf addresses ranging from 0 to 7. If the comparison results at depths 0, 1, and 2 are 1, 1, and 0, respectively, the resulting bitmap encodes the binary value 110, which corresponds to leaf address 6. 
In this way, the bitmap remaining in DRAM after all comparisons directly encodes, for each tree, the leaf address determined by the comparison results at each depth.

\noindent\textit{\textbf{Minimal CPU involvement.}} The \dtcr{3}{CPU} reads a single DRAM row to obtain all trees' leaf addresses, fetches the corresponding leaf values from DRAM, and sums them to produce the \dtcr{6}{final inference output (i.e., the predicted value)}. Both the comparisons at each node and the tree traversal based on the comparison results are performed entirely inside DRAM, with no intermediate data transferred off-chip. The CPU is only involved in the final step of aggregating leaf values and producing the \dtcr{6}{predicted value}.

\subsubsection{Performance and Energy Evaluation}
GBDT models are commonly deployed on resource-constrained environments such as edge devices or embedded systems~\cite{khataei2025treelut, tokuda2025df,zhang2021satellite}. To evaluate the performance under typical scenarios, we \dtcr{3}{use} a real, relatively low-power CPU system described in Table~\ref{tab:spec:gbdt}. We employ an optimized CPU implementation of CatBoost~\cite{prokhorenkova2018catboost,hancock2020catboost} by leveraging Arm NEON SIMD instructions~\cite{arm2024neon} along with multithreading to maximize performance on the baseline system. We estimate the CPU power consumption using the Xilinx Power Estimator~\cite{xilinx2022xpe}.
For a fair comparison, we evaluate the advantages of CatBoost inference using Clutch when PuD is integrated into this system. We evaluate PuD under this system configuration by assuming a parallelism of 64K columns $\times$ 16 banks, corresponding to a single DDR4 rank as described in Table~\ref{tab:spec:gbdt}. The number of chunks used for each bit-precision follows the configuration described in~\S\ref{sec:eva:per}.

\begin{table}[h]
\centering
\caption{Evaluated system configurations for GBDT}\label{tab:spec:gbdt}
\vspace{-1.0em}
\small
\setlength{\tabcolsep}{3pt}
\begin{tabular}{@{}ll@{}}
\toprule
\textbf{Main Memory} & 4 GB DDR4-2400~\cite{jedec2012ddr4}, single-channel, 64-bit, single rank; \\
\textbf{(\dtcr{3}{COTS} DRAM)}& 16 banks per rank, 8 KB row buffer per bank; \\
& Peak bandwidth: 19.2 GB/s \\
\midrule
\textbf{Processor} & Quad-core ARM Cortex-A53~\cite{arm2016cortexa53}, up to 1.5 GHz; \\
\textbf{(Real CPU)} & \textit{L1 Cache:} 32 kB D-cache, 32 kB I-cache; \\
& \textit{L2 Cache:} 1 MB shared, 64 B line size \\
\bottomrule
\end{tabular}
\end{table}

Since Clutch repeats vector–scalar comparisons for each feature, the performance of CatBoost inference on Clutch depends on the number of features. We therefore evaluate performance across datasets with varying feature counts.
We use large-scale, real-world tabular datasets listed in Table~\ref{tab:gbdt-data}, following prior work on GBDT benchmarking~\cite{prokhorenkova2018catboost}. We evaluate a multi-task inference composed of four independent tasks, each predicted by a GBDT model, following the standard multi-task scenario~\cite{iosipoi2022sketchboost, moshagen2010multitree}.
We vary the number of trees of the model across 512, 1024, and 2048, which we refer to as \textit{small}, \textit{medium}, and \textit{large} configurations, respectively. We also vary the depth of each tree across 8, 10, and 12.
Feature comparisons are evaluated at three different bit-precisions: 8-bit, 16-bit, and 32-bit. All leaf values are fixed at 16-bit precision. The default batch size is set to 1024.

\begin{table}[h]
  \centering
  \caption{Datasets for GBDT Inference.}
  \vspace{-1.0em}
  \small
  \label{tab:gbdt-data}
  \begin{tabular}{ccc}
    \toprule
    \textbf{Dataset} & \textbf{\# of Features} & \textbf{Size} \\
    \midrule
    airline~\cite{dataexpo2009airline}  & 13 & 115M  \\
    higgs~\cite{baldi2014higgs}     & 28 & 28M   \\
    covtype~\cite{blackard1999covertype}   & 54 & 581K  \\
    \bottomrule
  \end{tabular}
\end{table}

\begin{figure}[htbp]
    \centering
    \includegraphics[width=\columnwidth]{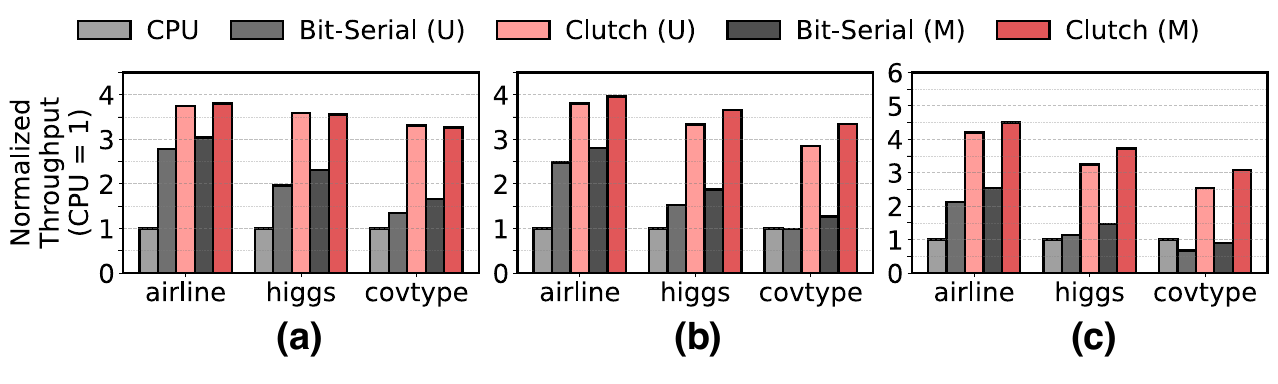}
    \vspace{-2.0em}
    \caption{Normalized throughput of GBDT inference at (a) 8-bit precision, (b) 16-bit precision, and (c) 32-bit precision.}
    \vspace{-0.5em}
    \label{fig:gbdt_thr}
\end{figure}

We first focus on the large model configuration with depth 10. Figure~\ref{fig:gbdt_thr} shows the throughput results for different datasets and bit-precisions. Clutch \dtcr{3}{provides} up to 4.5$\times$ (3.5$\times$ on average) speedup over CPU-based execution and up to 3.8$\times$ (2.2$\times$ on average) \dtcr{3}{speedup} over bit-serial PuD.
As the number of features increases, the execution time of both the bit-serial PuD and Clutch degrades due to the linear increase in the number of comparisons required to determine leaf addresses. Under such conditions, the advantage of Clutch becomes more pronounced compared to \dtcr{3}{the} bit-serial PuD approach. For example, \dtcr{3}{for the} 32-bit inference \dtcr{6}{with} the \textit{covtype} dataset, the bit-serial PuD method performs worse than the CPU baseline, whereas Clutch (M) provides a 3.1$\times$ speedup \dtcr{3}{over the CPU}. 

\begin{figure}[H]
    \centering
    \includegraphics[width=\columnwidth]{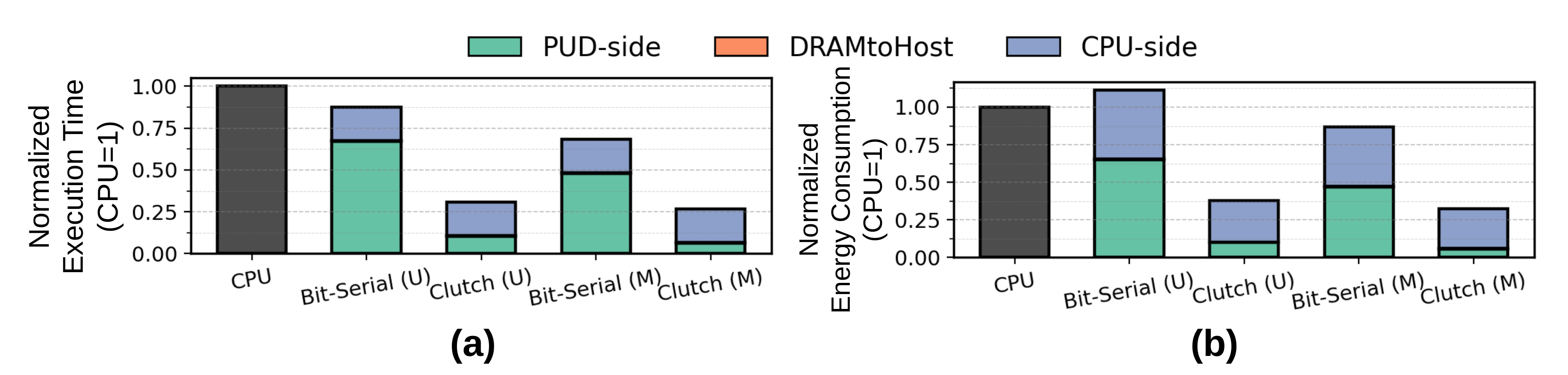}
    \vspace{-2.0em}
    \caption{Breakdown of (a) execution time and (b) energy consumption of GBDT inference.}
    \label{fig:gbdt_bd}
\end{figure}

Figure~\ref{fig:gbdt_bd} presents the breakdown of execution time and energy consumption for the 32-bit \textit{higgs} dataset. For the PuD implementations, the breakdown consists of three components: 1) PuD execution that performs in-DRAM leaf-address retrieval (\textit{PuD-side}), 2) transfer of the leaf addresses from DRAM to the host (\textit{DRAMtoHost}), and 3) CPU-side summation of leaf values (\textit{CPU-side}).

In Bit-Serial (M), the PuD-side component is responsible for 70.2\% of the total execution time, while CPU-side component accounts for 29.2\% and DRAMtoHost for 0.6\% of the total execution time. In Clutch (M), the PuD-side component reduces to 23.9\% and DRAMtoHost accounts for 1.5\%, shifting the dominant bottleneck to the CPU-side at 74.6\%. This is because Clutch reduces the PuD-side operation cost by 7.0$\times$ on average compared to bit-serial PuD.

A similar trend is observed in energy consumption. In Bit-Serial (M), the PuD-side accounts for 54.2\% of total system energy, while CPU-side accounts for 45.6\% and DRAMtoHost accounts for 0.3\%. In Clutch (M), the PuD-side share reduces to 18.0\%, while DRAMtoHost accounts for 0.7\%, with the CPU-side becoming the dominant contributor at 81.3\%. Overall, Clutch provides a 2.9$\times$ energy efficiency improvement over the CPU and 2.8$\times$ over bit-serial PuD on average.

\begin{figure}[H]
    \centering
    \includegraphics[width=0.80\columnwidth]{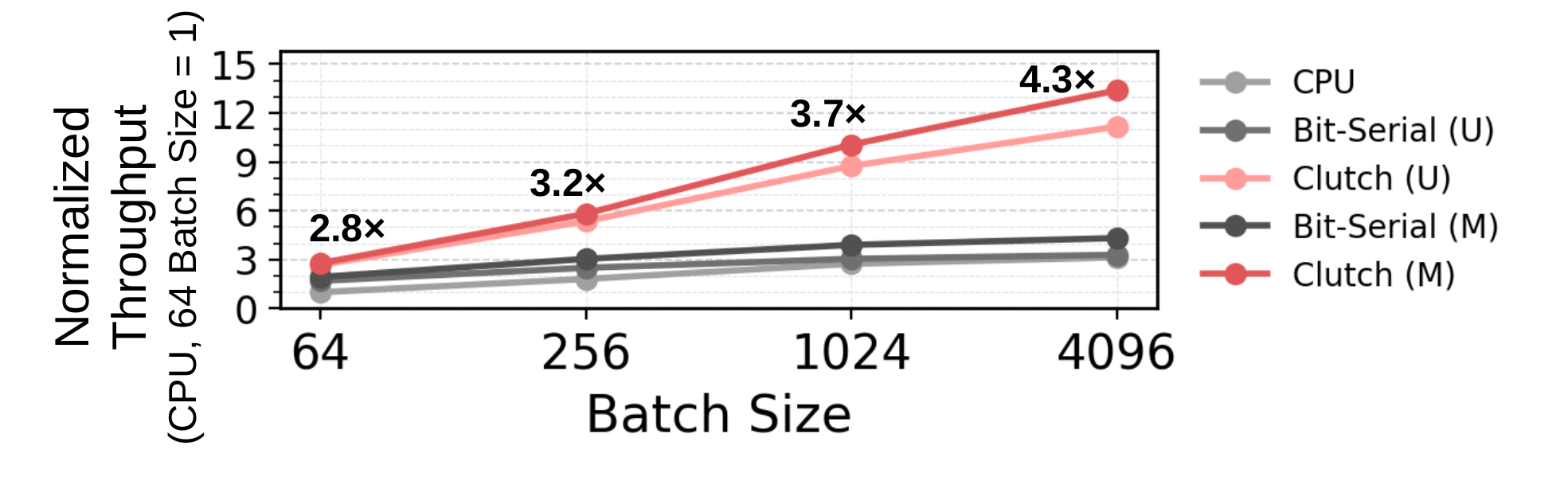}
    \vspace{-1em}
    \caption{Sensitivity of GBDT inference throughput to batch size.}
    \vspace{-1.0em}
    \label{fig:gbdt_batch}
\end{figure}

Figure~\ref{fig:gbdt_batch} shows the sensitivity to batch size for the 32-bit \textit{higgs} dataset. Throughput is normalized to the CPU baseline with a batch size of 64. The numbers annotated next to each Clutch (M) data point indicate the throughput improvement of Clutch (M) over the CPU at the corresponding batch size. As batch size increases, CPU L1 and L2 data cache hit rate improves during the CPU-based accumulation of leaf values, resulting in faster execution. Consequently, the comparison operations become more dominant in overall execution time, making the throughput advantage of Clutch over both the CPU and bit-serial PuD more pronounced. At a batch size of 4096, Clutch (M) provides 4.3$\times$ speedup over the CPU and 3.3$\times$ over Bit-Serial (M).

\begin{figure}[htbp]
    \centering
    \includegraphics[width=\columnwidth]{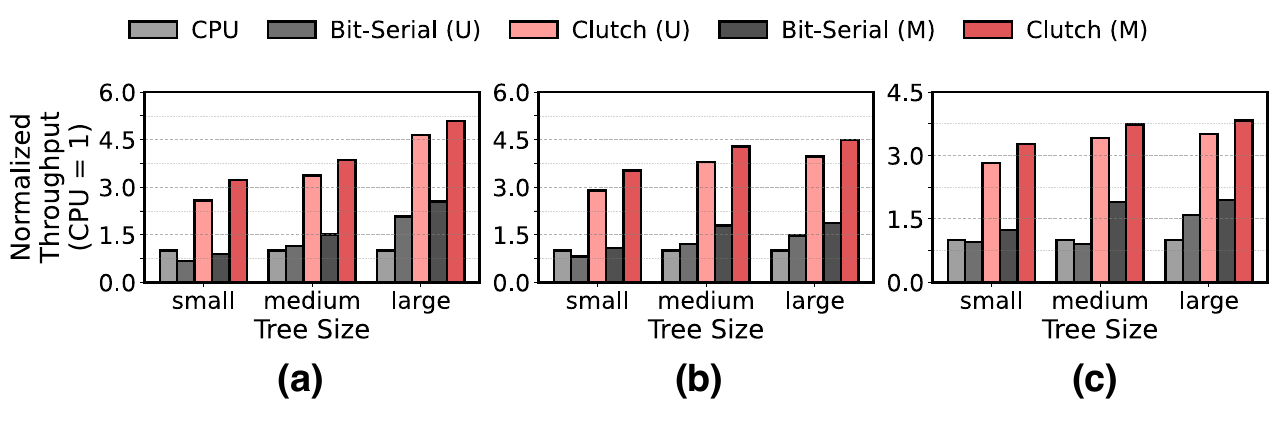}
    \vspace{-1.8em}
    \caption{Sensitivity of GBDT inference throughput to model size at (a) 8-bit precision, (b) 16-bit precision, and (c) 32-bit precision.}
    \vspace{-0.5em}
    \label{fig:gbdt_config}
\end{figure}

To evaluate the generality of Clutch across various model configurations, we use the \textit{higgs} dataset and test \dtcr{3}{various} combinations of tree sizes and depths. Figure~\ref{fig:gbdt_config} demonstrates that Clutch (M) consistently outperforms other methods across different model settings, \dtcr{3}{providing} up to 5.1$\times$ (3.9$\times$ on average) speedup over the CPU and up to 3.6$\times$ (2.5$\times$ on average) over Bit-Serial (M). In particular, when the tree size is \textit{small}, bit-serial PuD suffers significant throughput loss due to insufficient utilization of parallelism. In contrast, Clutch maintains high performance even in such cases thanks to its faster execution of \dtcr{3}{in-DRAM comparison}, \dtcr{3}{providing} 2.9$\times$ higher throughput than the CPU on average across Clutch (U) and Clutch (M).

\subsubsection{Data Conversion and  Memory Footprint Overheads}
\begin{figure}[htbp]
    \centering
    \vspace{-0.5em}
    \includegraphics[width=\columnwidth]{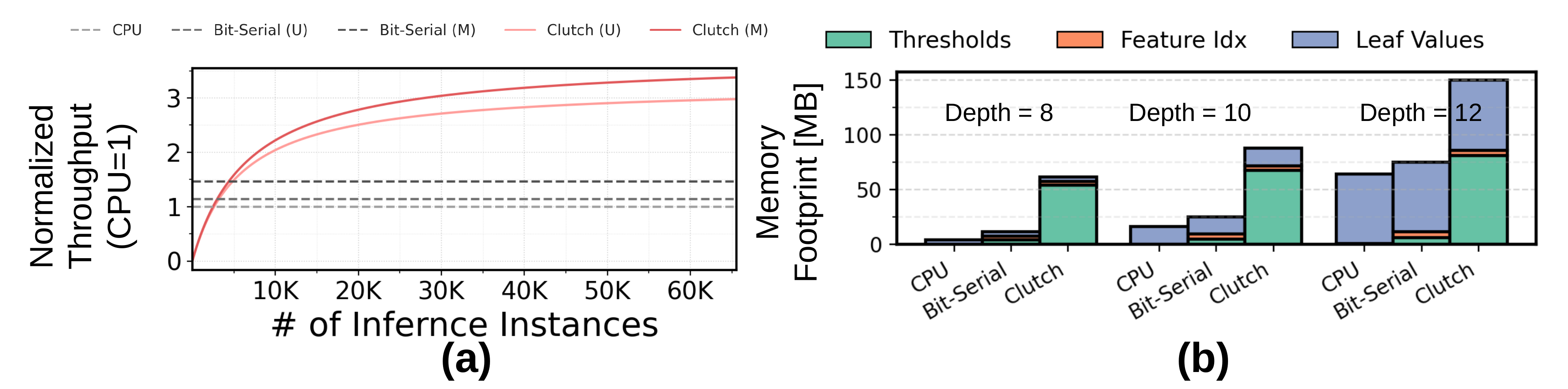}
    \vspace{-1.8em}
    \caption{Clutch overhead analyses. (a) Effective GBDT inference throughput including the data conversion overhead. (b) Memory footprint.}
    \vspace{-0.5em}
    \label{fig:gbdt_overhead}
\end{figure}

We quantitatively evaluate the overhead introduced by the Clutch encoding scheme in terms of data conversion time and memory footprint during CatBoost inference.
~\dtcr{3}{Clutch requires a one-time conversion of the vector data from binary to chunked temporal coding before inference starts. One way to eliminate this overhead is to pre-convert the static vector data offline and store the converted representation in DRAM before execution. Another approach is to perform the conversion once at runtime and amortize its cost over subsequent \dtcr{6}{inference instances (i.e., input data points)}. \dtcr{6}{As more instances are processed, the per-instance overhead decreases.} Figure~\ref{fig:gbdt_overhead}a shows the effective throughput (i.e., the throughput measured from the start of conversion through inference), including the initial conversion overhead, as a function of the number of inference instances processed. \dtcr{6}{The point at which Clutch's effective throughput curve crosses the CPU baseline is approximately 5K inference instances}, demonstrating that the conversion overhead is quickly amortized in practical workloads.}

Figure~\ref{fig:gbdt_overhead}b compares the memory footprint when the tree size is \textit{large} and using 32-bit precision. \dtcr{6}{Since the memory footprint is identical for Bit-Serial (U) and Bit-Serial (M), and likewise for Clutch (U) and Clutch (M), Figure~\ref{fig:gbdt_overhead}b shows one entry each for Bit-Serial and Clutch.}
\dtcr{6}{In Figure~\ref{fig:gbdt_overhead}b, \textit{Feature Idx} refers to the memory footprint of the feature index stored for each node, which corresponds to the one-hot feature masks in Clutch's layout.} For deeper trees (depth = 12), leaf values dominate the total memory footprint, and Clutch introduces an additional encoding overhead of about 85~MB compared to the baseline model. For shallow trees (depth = 8), the relative overhead of Clutch becomes larger because the baseline model itself is smaller; however, the absolute footprint of the Clutch-encoded model remains modest at around 60~MB in total, which is \dtcr{6}{within 1--2\% of the 4--8~GB DRAM capacity} typically available on embedded devices.

\subsection{Predicate Evaluation}\label{sec:app:que}
Predicate evaluation is a fundamental operation in data-centric applications, appearing in database queries~\cite{willhalm2009simd, graefe2011modern, li2013bitweaving, farber2012sap, grund2010hyrise, idreos2012monetdb, kemper2012hyper, lahiri2015oracle}, scientific computing, and image processing~\cite{ergin2017dynamic, zhong2022using, santitissadeekorn2020approximate, abdusalomov2020automatic, Li2021_HTMaskRCNN}. It applies numerical or categorical predicates (e.g., inequalities or category-membership tests) and produces the bitmap indicating whether each element satisfies the predicate. This bitmap enables efficient filtering and reduces the workload for subsequent computations.
In in-memory databases, predicate evaluation is typically implemented via column scans in the \texttt{WHERE} clause, playing a key role in early-stage query processing and lowering the cost of operations such as aggregation~\cite{willhalm2009simd, graefe2011modern, li2013bitweaving, farber2012sap, grund2010hyrise, idreos2012monetdb, kemper2012hyper, lahiri2015oracle}. We evaluate Clutch on predicate evaluations from in-memory database workloads.

In the PuD implementation, each DRAM column corresponds to one record, and all feature values of that record are placed vertically within the same column in the same subarray~\cite{seshadri2017ambit}. In Clutch's implementation, each feature vector is encoded using chunked temporal coding. Because the Clutch algorithm described in Algorithm~\ref{alg:conquer} supports only the $<$ comparison, other comparison operators are derived as follows. The $\le$ operator is realized by decrementing the scalar value by one, since $a \le b$ is equivalent to $(a - 1) < b$ for integers. The $>$ and $\ge$ operators are obtained by negating the results of $\le$ and $<$, respectively. On Modified PuD, this negation is performed using bulk bitwise \myNOT{} operations, while on Unmodified PuD, which lacks the native \myNOT{}, the complement of each feature value is additionally stored to obtain the negated result; ~\dtcr{3}{this is analogous to how the bit-serial PuD on Unmodified DRAM also maintains complemented values to support all comparison operators~\cite{gao2019computedram, kubo2025mvdram}.} Finally, the $==$ operator is computed as the bitwise \texttt{AND} of $\le$ and $\ge$. In this way, Clutch-based PuD supports all five comparison operators ($<$, $\le$, $>$, $\ge$, and $==$).

For each predicate, PuD produces a bitmap directly in DRAM, and bitwise \texttt{AND} or \texttt{OR} operations are then performed within the same subarray to produce the final bitmap corresponding to the \texttt{WHERE} clause. This resulting bitmap is then loaded by the processor, which performs subsequent operations such as \texttt{COUNT} or \texttt{AVERAGE}.

\subsubsection{Evaluation}
We evaluate the \dtcr{3}{effect} of Clutch-based acceleration of predicate evaluation. Since Clutch's performance is sensitive to the bit-precision of feature values, and existing general-purpose query evaluation benchmarks do not allow flexible control over bit-precision, we develop our own benchmark, as done in prior work~\cite{seshadri2017ambit,li2013bitweaving, wang2018rc, xin2021sam, fujiki2023mvc}. 
Our benchmark captures a broad range of query patterns, from common to more intricate cases, as summarized in Table~\ref{tab:que}.
Each dataset consists of eight features sampled from a uniform distribution. We use three levels of bit-precision (8-bit, 16-bit, and 32-bit) and three different data sizes measured in the total number of feature values (\textit{small table}: 64M, \textit{medium table}: 256M, and \textit{large table}: 1G), corresponding to 8M, 32M, and 128M records, respectively.

\begin{table}[h]
\centering
\caption{Benchmark queries}\label{tab:que}
\vspace{-1em}
\small
\begin{tabular}{@{}ll@{}}
\toprule
\textbf{Q1} & WHERE $x_0 < f_i < x_1$  \\
\midrule
\textbf{Q2} & WHERE ($x_0 < f_i < x_1$ AND  $y_0 < f_j < y_1$)\\
\midrule
\textbf{Q3} & COUNT (WHERE ($x_0 < f_i < x_1$ OR $y_0 < f_j < y_1$))\\
\midrule
\textbf{Q4} & AVERAGE($f_k$) FROM (WHERE $x_0 < f_i < x_1$ AND $y_0 < f_j < y_1$)\\
\midrule
\textbf{Q5} & WITH avg\_val = AVERAGE($f_k$) \\
   & \quad \quad \quad \quad WHERE $(x_0 < f_i < x_1 \;\text{OR}\; y_0 < f_j < y_1)$ \\
   & COUNT (WHERE avg\_val $< f_\ell < 2 \cdot$ avg\_val )\\
\bottomrule
\end{tabular}
\end{table}

We evaluate Clutch's performance on 1) a desktop-class CPU system as described in Table~\ref{tab:spec:intel} and 2) a server-class GPU system as summarized in Table~\ref{tab:spec:gpu}. Following prior work~\cite{liu2025optipim}, we project the throughput of PuD when integrated into the A100's HBM2 memory, assuming that the HBM2 subarray structure can support PuD operations as in DDR4. In this analysis, we assume a per-stack parallelism of 2KB columns $\times$ 16 banks $\times$ 8 channels.

\begin{table}[h]
\centering
\caption{Evaluated system configurations for predicate evaluation on GPU}\label{tab:spec:gpu}
\vspace{-0.5em}
\small
\begin{tabular}{@{}ll@{}}
\toprule
\textbf{Main Memory} & 5 stacks of HBM2, total 40 GB; \\
\textbf{(COTS DRAM)} & Peak bandwidth: 1555 GB/s; \\
\midrule
\textbf{Processor} & NVIDIA A100 Tensor Core GPU (PCIe), up to 1.59 GHz; \\
\textbf{(Real GPU)} & \textit{SMs:} 108, \textit{CUDA Cores:} 6912; \\
\bottomrule
\end{tabular}
\end{table}

We compare the PuD approach against BitWeaving-V~\cite{li2013bitweaving}, a state-of-the-art implementation for predicate evaluation on CPU. BitWeaving-V stores data in a transposed, bit-sliced layout optimized for predicate evaluation, and we confirm that it provides higher performance than conventional layouts for all tested bit-precisions on both CPU and GPU. 
Since both the CPU baseline using BitWeaving-V and PuD operate on transposed layouts that are not suited for value retrieval, all platforms (CPU, GPU, and PuD) also maintain a copy of the database in a conventional layout to efficiently perform post-processing operations such as \texttt{AVERAGE}.
Clutch uses the following number of chunks: 8-bit: 2 chunks, 16-bit: 4 chunks, 32-bit (Modified PuD): 8 chunks, and 32-bit (Unmodified PuD): 12 chunks, so that all feature vectors fit within a subarray.~\footnote{On Unmodified PuD, each feature value and its complement must both be stored to support all comparison operators. Depending on the configuration, a larger number of chunks can be required to fit all features within the row budget of a single subarray.}

\begin{figure}[htbp]
    \centering
    \includegraphics[width=\columnwidth]{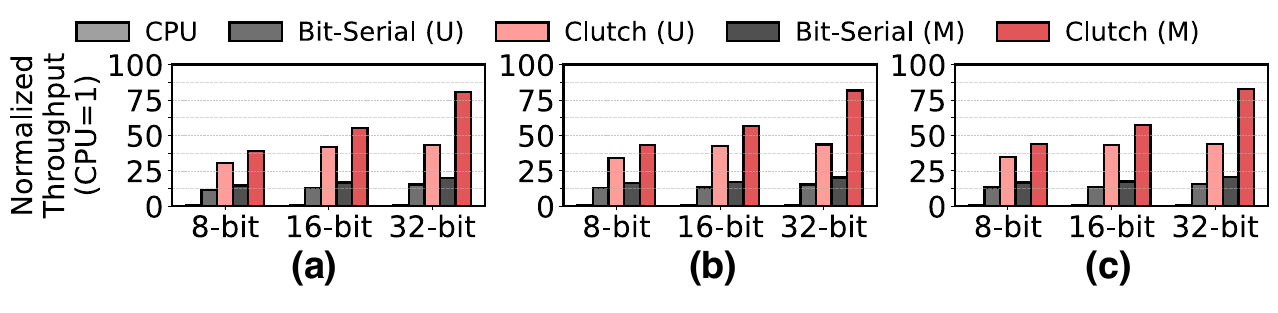}
    \vspace{-2.0em}
    \caption{Normalized throughput of Q2 at (a) small table, (b) medium table, and (c) large table.}
    \label{fig:ts_thr}
\end{figure}

We first focus on the evaluation of Q2 on the CPU-based system. Figure~\ref{fig:ts_thr} shows the normalized throughput relative to CPU for three table sizes and three bit-precisions. Clutch \dtcr{3}{provides} up to 83$\times$ (50$\times$ on average) speedup over the CPU. This improvement is due to the ability of PuD to reduce data transfers for both comparison operations and bitmap reductions. Compared to bit-serial PuD, Clutch \dtcr{3}{provides} up to 4.0$\times$ (3.1$\times$ on average) speedup.

\begin{figure}[htbp]
    \centering
    \includegraphics[width=\columnwidth]{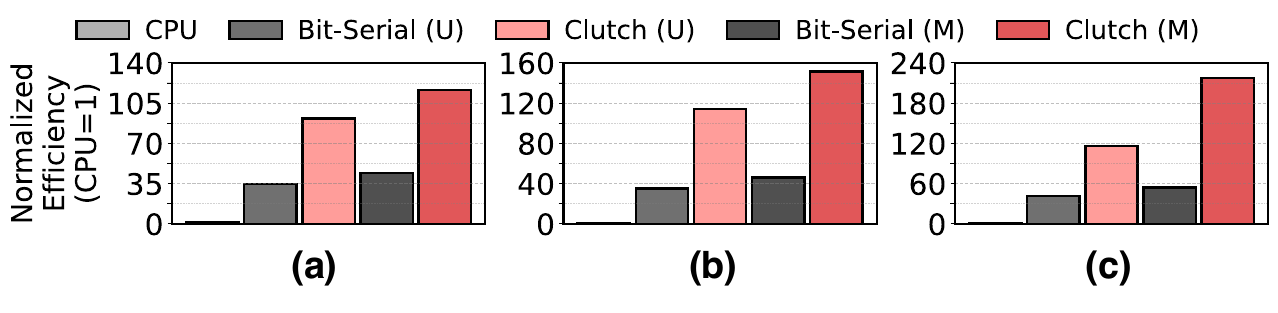}
    \vspace{-2.0em}
    \caption{Normalized energy efficiency of Q2 at (a) 8-bit precision, (b) 16-bit precision, and (c) 32-bit precision.}
    \label{fig:ts_ene}
\end{figure}

Figure~\ref{fig:ts_ene} presents the energy efficiency normalized to CPU for the large table configuration. Clutch provides up to 218$\times$ (135$\times$ on average) energy efficiency improvement over the CPU and up to 4.2$\times$ (3.1$\times$ on average) over bit-serial PuD.

\begin{figure}[htbp]
    \centering
    \includegraphics[width=\columnwidth]{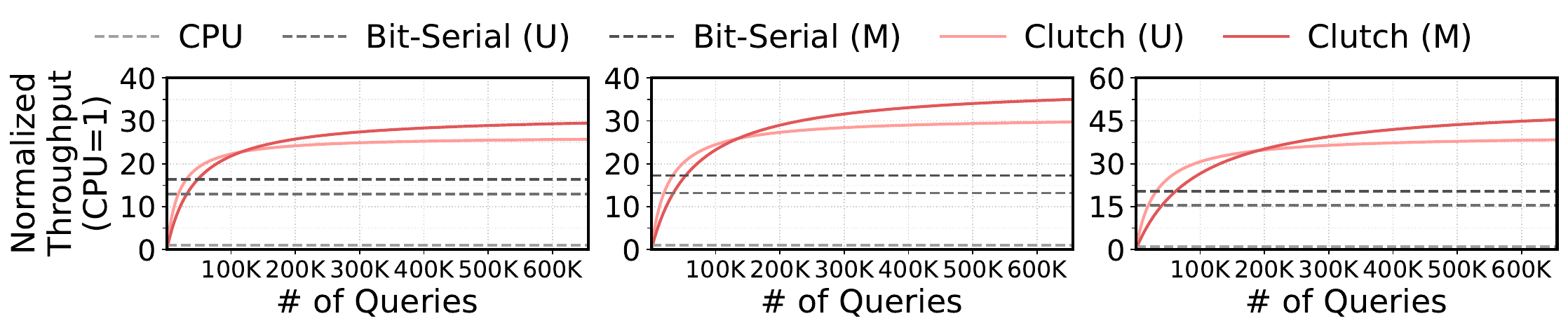}
    \vspace{-1.8em}
    \caption{Effective throughput of Q2 including the data
conversion overhead at (a) 8-bit precision, (b) 16-bit precision, and (c) 32-bit precision.}
    \vspace{-0.5em}
    \label{fig:ts_conv}
\end{figure}

We quantitatively evaluate the overhead of data format conversion for predicate evaluation at 32-bit precision, following the same methodology as the GBDT evaluation (Figure~\ref{fig:gbdt_overhead}a). As with GBDT inference, Clutch requires a conversion of the vector data from binary to chunked temporal coding before execution begins. Figure~\ref{fig:ts_conv} shows the effective throughput, including this initial conversion overhead, as a function of the number of queries processed. Since the conversion cost is fixed, it is amortized over subsequent queries (i.e., the per-query overhead decreases as more queries are processed). Clutch surpasses the CPU throughput after processing approximately 1.5K--1.9K queries, and outperforms the bit-serial PuD after processing approximately 48K--61K queries. If this amortization is insufficient, the runtime conversion overhead can be eliminated entirely by pre-converting the data offline and storing it in DRAM before execution.

\begin{figure}[htbp]
    \centering
    \includegraphics[width=\columnwidth]{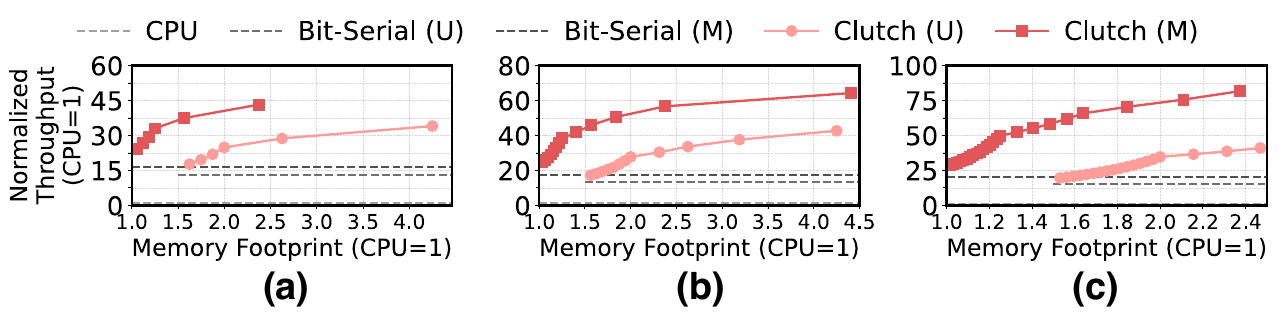}
    \vspace{-2.0em}
    \caption{Tradeoff between throughput and memory footprint of Q2 at (a) 8-bit precision, (b) 16-bit precision, and (c) 32-bit precision.}
    \vspace{-0.5em}
    \label{fig:ts_cap}
\end{figure}

\dtcr{3}{Clutch provides a flexible tradeoff between throughput and memory footprint by adjusting the chunk count. Figure~\ref{fig:ts_cap} illustrates this tradeoff space between memory footprint and the throughput of Q2. Within 1.5$\times$ the memory footprint of the CPU execution, Clutch (M) provides up to 58$\times$ (44$\times$ on average) higher throughput compared to the CPU, and within 2.0$\times$, up to 70$\times$ (53$\times$ on average). Compared to Bit-Serial (M), Clutch (M) provides up to 2.9$\times$ (2.4$\times$ on average) higher throughput within 1.5$\times$ the memory footprint of the bit-serial PuD, and up to 3.5$\times$ (2.9$\times$ on average) within 2.0$\times$.}

\begin{figure}[htbp]
    \centering
    \includegraphics[width=\columnwidth]{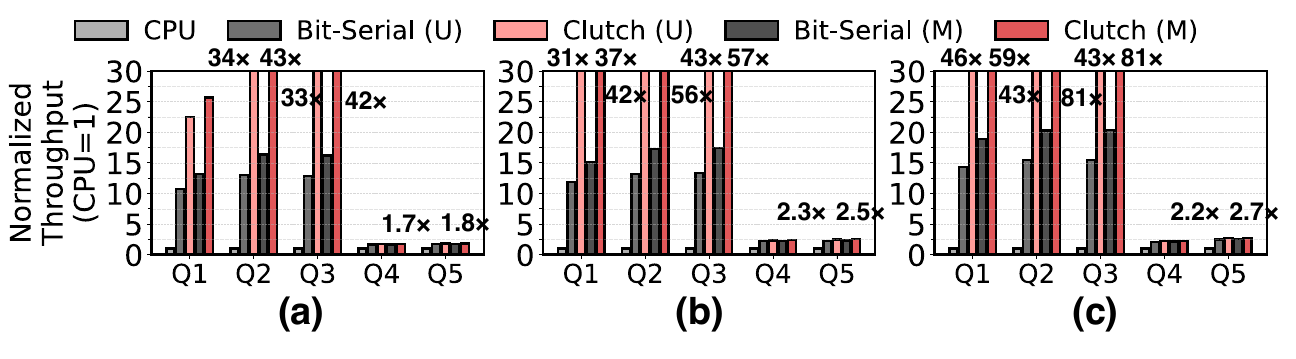}
    \vspace{-2.2em}
    \caption{Normalized throughput on CPU-based system at (a) 8-bit precision, (b) 16-bit precision, and (c) 32-bit precision.}
    \label{fig:ts_q5_cpu}
\end{figure}

\begin{figure}[htbp]
    \centering
    \includegraphics[width=\columnwidth]{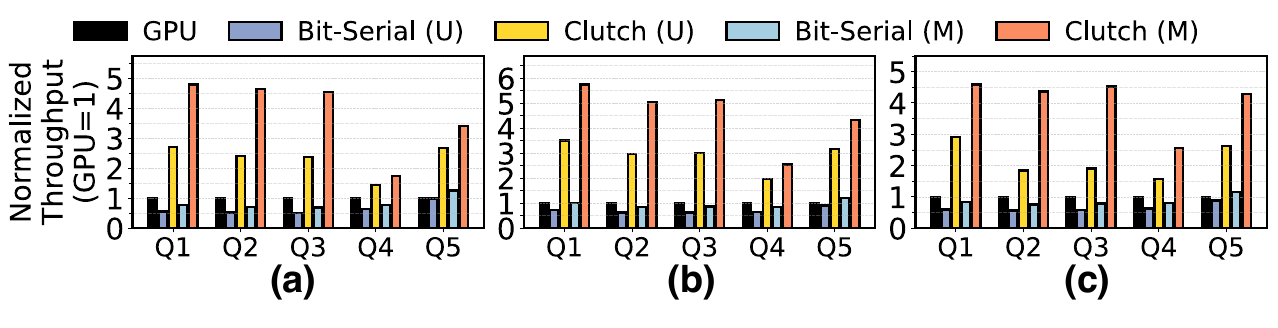}
    \vspace{-2.2em}
    \caption{Normalized throughput on GPU-based system at (a) 8-bit precision, (b) 16-bit precision, and (c) 32-bit precision.}
    \label{fig:ts_q5_gpu}
\end{figure}

\begin{figure}[htbp]
    \centering
    \includegraphics[width=0.80\columnwidth]{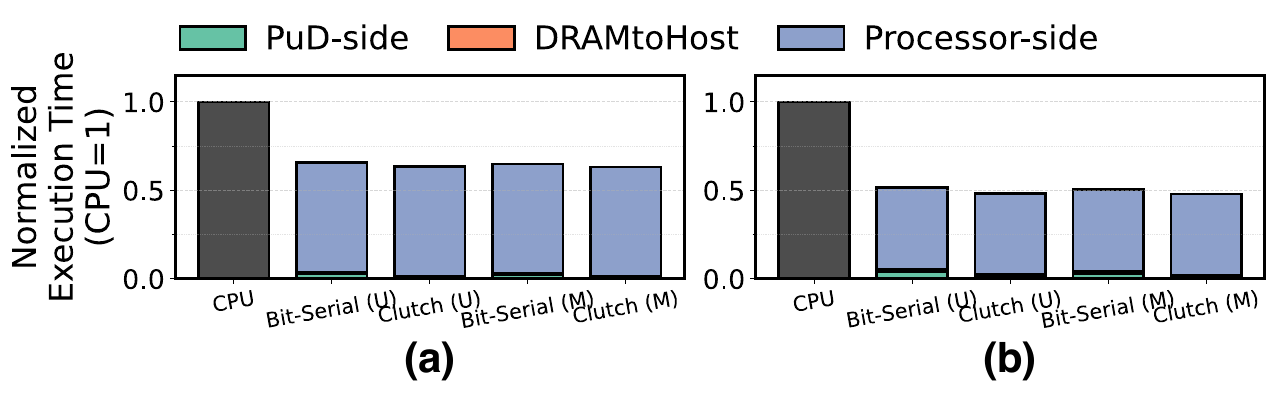}
    \vspace{-1.5em}
    \caption{Breakdown of execution time on CPU-based system at (a) Q4 and (b) Q5.}
    \label{fig:ts_bd_cpu}
\end{figure}

\begin{figure}[htbp]
    \vspace{-0.5em}
    \centering
    \includegraphics[width=0.80\columnwidth]{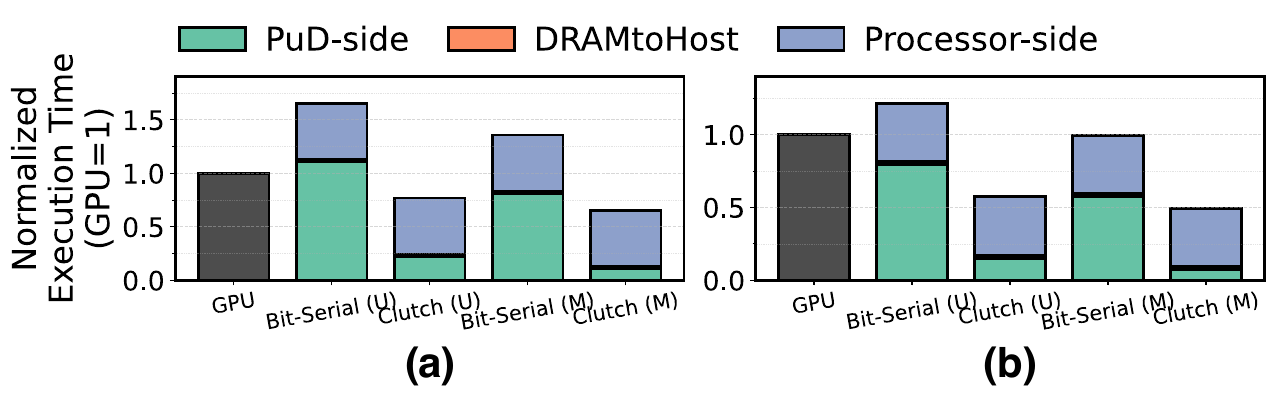}
    \vspace{-1.5em}
    \caption{Breakdown of execution time on GPU-based system at (a) Q4 and (b) Q5.}
    \vspace{-0.5em}
    \label{fig:ts_bd_gpu}
\end{figure}

Figure~\ref{fig:ts_q5_cpu}~\footnote{\dtcr{3}{In Figure~\ref{fig:ts_q5_cpu}, the numbers annotated near each bar indicate the speedup over the CPU baseline. For Q4 and Q5, the annotated numbers correspond to Clutch (M).}} and Figure~\ref{fig:ts_q5_gpu} show the throughput for \dtcr{3}{all} five queries listed in Table~\ref{tab:que} executed on the medium table using the CPU system (Table~\ref{tab:spec:intel}) and the GPU system (Table~\ref{tab:spec:gpu}), respectively.
On the CPU system, Clutch \dtcr{3}{provides} up to 81$\times$ (28$\times$ on average) higher throughput than the CPU and up to 4.0$\times$ (2.2$\times$ on average) over the bit-serial PuD implementation for Q1 through Q5. For Q4 and Q5, where post-processing operations dominate, Clutch \dtcr{3}{provides} up to 2.7$\times$ (2.1$\times$ on average) speedup over the CPU baseline; however, its additional speedup over the bit-serial PuD implementation is limited to up to 1.08$\times$ (1.05$\times$ on average).
To understand this trend, Figure~\ref{fig:ts_bd_cpu} and Figure~\ref{fig:ts_bd_gpu} show the breakdown of execution time at 8-bit precision on the CPU-based and GPU-based systems, respectively. \dtcr{3}{On the CPU-based system, in the Bit-Serial (M) implementation, post-bitmap processing on the processor side accounts for 96\% of the total execution time for Q4 and 92\% for Q5. Since Clutch accelerates only the PuD-side execution, the overall throughput improvement is inherently limited for these queries.}

In contrast, on the GPU system, the bit-serial PuD implementation fails to outperform processor execution in most cases, whereas Clutch provides up to 5.7$\times$ (3.3$\times$ on average) speedup over GPU-based execution and up to 6.6$\times$ (4.3$\times$ on average) over bit-serial PuD. 
This is because, \dtcr{3}{in} HBM2, \dtcr{3}{the ratio of column-level parallelism to memory bandwidth is smaller than that} on DDR4, making \dtcr{3}{the bit-serial} in-DRAM comparisons a more prominent bottleneck. 
\dtcr{3}{As a result, even for Q4 and Q5, the PuD-side comparison accounts for 63\% and 61\% of the total execution time on average in the Bit-Serial (M) implementation}, as shown in Figure~\ref{fig:ts_bd_gpu}.
\dtcr{3}{By accelerating this dominant component, Clutch provides end-to-end throughput gains on the GPU system.}
This highlights Clutch's advantage in efficiently handling comparison-intensive workloads even under limited column-level parallelism, helping to unlock the full performance potential of future PuD architectures integrated with HBM.

\section{Discussion}
\subsection{Limitations of Clutch}\label{sec:dis:lim}
Clutch does not always deliver significant throughput improvements compared to \dtcr{3}{CPU}-based execution or bit-serial PuD implementations. This section discusses the primary scenarios where Clutch \dtcr{3}{can} be less effective.

\subsubsection{Comparison is Not the Dominant Bottleneck}
Clutch can fail to provide high throughput when comparison operations are not the dominant performance bottleneck. For instance, on CPU-based systems, the execution of Q3, Q4, and Q5 with bit-serial PuD shows limited improvement due to the dominant cost of post-comparison processing (see Figure~\ref{fig:ts_bd_cpu}).
To address this limitation, even in such cases, offloading non-comparison operations to dedicated hardware integrated within the memory controller can enable applications to benefit from Clutch’s high-speed comparison \dtcr{3}{operations}. Clutch has the potential to be effectively combined with other specialized processing units \dtcr{3}{including Processing-near-Memory techniques~\cite{ahn2015scalable, ahn2015pim, akin2015datareorg, akin2014hamlet, alves2016largevector, alves2015vectorops, alves2015savingmovements, asghari2016singleisa, asghari2016chameleon, azarkhish2016logicbase, azarkhish2016case, azarkhish2018neurostream, babarinsa2015jafar, besta2021sisa, boroumand2020practical, boroumand2021googleedge, boroumand2021mitigatingedge, boroumand2018googleworkloads, boroumand2017lazypimcal, boroumand2021polynesiaarxiv, boroumand2022polynesia, boroumand2019conda, boroumand2017lazypimarxiv, cali2020genasm, cho2020mcdramv2, dai2018graphh, delima2018designspace, denzler2021casper, devaux2019truepim, drumond2017mondrian, elliott1999computationalram, farmahini2014drama, farmahini2015nda, fernandez2020natsa, gao2016hrl, gao2017tetris, ghiasi2022genstore, giannoula2022sparsep, giannoula2021syncron, gokhale1995terasys, gomezluna2021benchmarkingcut, gomezluna2022benchmarkingaccess, gomezluna2021benchmarkingarxiv, gu2016hardwaresecurity, gu2020ipim, guo2014memoryside, hadidi2017cairo, hall1999diva, hsieh2016tom, hsieh2016pointerchasing, huang2019activerouting, huang2020heterogeneouspim, jang2019charon, ke2021axdimm, kersey2017lightweightsimt, kim2016neurocube, kim2018grimfilter, kim2017grimfilterarxiv, kwon2021fimdram, lee2015bssync, lee2021pimproduct, lee2022gddr6aim, li2019pims, lim2017tep, liu2018heterogeneoustraining, lockerman2020livia, matam2019graphssd, nai2017graphpim, nair2015activememorycube, niu2022processnearmemory, oliveira2017nim, patterson1997intelligentram, pattnaik2016gpupim, pugsley2014ndc, santos2018processing3d, santos2017operandsize, shin2018mcdram, singh2019napel, singh2020nero, sun2021abcdimm, tsai2018adaptivescheduling, xi2015beyondwall, zhang2014top, zhang2018graphp, zhu2013spgemm3d, zhuo2019graphq}} to enhance overall system performance.

\subsubsection{Limited Benefit with Small Working Sets}
As discussed in~\S\ref{sec:mot}, one key source of PuD’s performance gain is its ability to reduce off-chip data movement. When the vector working set fits entirely within the processor’s cache hierarchy, however, this benefit largely disappears: data can be reused from the cache, and a sufficiently powerful processor can, in principle, outperform PuD. In such small–working set scenarios (e.g., inference with small GBDT models or predicate evaluation over small tables on server-class processors) conventional processor-centric execution can be the more appropriate choice.

Even when the working set fits in the CPU cache, Clutch can still exceed the processor’s throughput in some cases. Clutch offers very high comparison throughput inside DRAM (e.g., \dtcr{3}{providing} 4.9 TOPS for 16-bit comparison  under the configuration of Table~\ref{tab:spec:intel}). Moreover, it produces 1-bit-per-element bitmaps, reducing off-chip data movement, and these bitmaps can be combined using high-throughput in-DRAM bitwise operations. 

\subsubsection{Limited Benefit with Non-Static Vector Data}
Although we quantitatively showed that the overhead of data conversion for Clutch can be amortized when the converted vector data is reused repeatedly during application execution (see Figure~\ref{fig:gbdt_overhead}a and Figure~\ref{fig:ts_conv}), this amortization benefit diminishes when the vector data is not static. If the vector is frequently updated throughout the application, Clutch cannot deliver \dtcr{3}{as} high performance. This is because its specialized encoding scheme incurs a high cost during vector updates. This motivates future work on extending Clutch to efficiently support frequently updated vectors.

\subsubsection{When Memory Capacity Headroom Is Absent}
Clutch's specialized encoding incurs memory capacity overhead. When sufficient headroom is unavailable, \dtcr{3}{this overhead can cause some data to be paged out, undermining overall performance gains.} In such cases, the system should fall back to the bit-serial PuD. However, Clutch can adjust its chunk count to operate within a given capacity budget and still deliver substantial speedups (e.g., within 1.5$\times$ the baseline memory footprint, Clutch provides up to 58$\times$ higher throughput over the CPU for Q2).

\subsection{Synergy Between Clutch Algorithm and PuD Execution}
\dtcr{2}{Clutch's temporal-coding algorithm is designed to be general and can, in principle, be executed on a conventional processor.
However, PuD execution uniquely amplifies its benefit.
In a processor-based execution, the processor must read every per-chunk bitmap from DRAM to combine them into the final comparison result, generating data transfer proportional to the number of chunks.
PuD eliminates this overhead by performing all per-chunk lookups and their combination entirely within DRAM, so that only the final result bitmap needs to be transferred to the processor.
Consequently, Clutch-PuD reduces data transfer by $5\times$ for 32-bit kernel-level comparisons, and by up to $140\times$ (Figure~\ref{fig:gbdt_thr}) and $17\times$ (Figure~\ref{fig:ts_thr}) at the application level compared to executing the same Clutch algorithm on the \dtcr{3}{CPU}. These results show the strong affinity between \dtcr{3}{the} Clutch algorithm and the PuD paradigm.}

\subsection{System Integration}\label{sec:dis:sys}
Clutch requires the host processor to dynamically issue PuD operations based on the scalar value the host processor holds. Clutch follows the host-driven system integration model introduced by prior work~\cite{de2026count2multiply,kubo2025mvdram}, where the host processor dynamically constructs a $\mu$Program based on the scalar value and dispatches it to the memory-controller-side control unit for execution.

From \dtcr{3}{the PuD execution framework} originally introduced by SIMDRAM~\cite{hajinazar2021simdram}, Clutch adopts a subset of its mechanisms: (i)~a small ISA extension that provides an invocation surface to trigger PuD operations on selected memory regions, (ii)~OS support to manage PIM objects, and (iii)~a lightweight control unit associated with the memory controller that executes the $\mu$Program and issues the corresponding DRAM commands with the timing parameters required for PuD operations such as \RowCopy{} and \MAJ{3}.

Unlike SIMDRAM, Clutch does not require \dtcr{3}{a} hardware data-transposition unit that converts data between horizontal and vertical layouts at runtime. In Clutch, the scalar value is initialized within DRAM through \RowCopy{} from constant rows, eliminating the need for runtime layout conversion and its associated hardware complexity. As a result, Clutch can be realized with modest extensions to the memory controller and \dtcr{3}{small} ISA/OS \dtcr{3}{modifications}.

\subsection{Outlook for Future DRAM and PuD’s Potential}
PuD’s advantage can be understood by a balance: (i) how much internal column-level parallelism a DRAM module can expose per ACT, versus (ii) \dtcr{3}{the off-chip bandwidth available to processor-based execution}. As newer DRAM technologies \dtcr{3}{increase} off-chip bandwidth, holding internal parallelism fixed would shrink PuD’s relative edge.
However, PuD-oriented optimizations (e.g., increasing column-level parallelism and aggressively exploiting subarray-level parallelism (SALP)~\cite{kim2012case, oliveira2025proteus}) can increase PuD computation throughput. With these enhancements, PuD’s advantage can be preserved \dtcr{3}{and} even amplified across DRAM generations.

\section{Related Work}

\subsubsection*{Other PuD architectures}
Although various PuD architectures have been proposed~\cite{kim2012case, seshadri2013rowclone, seshadri2015fast, seshadri2016buddy, seshadri2016processing, li2017drisa, deng2018dracc, seshadri2019dram,deng2019lacc, wang2020figaro, xin2020elp2im, hajinazar2021simdram,li2018scope,thakkar2023low,
ferreira2022pluto, wu2022dram, deng2023dram, oliveira2024mimdram, shivanandamurthy2021atria, afifi2024artemis, oliveira2025proteus}, Clutch differs from these proposals in two key aspects. First, by introducing a data representation tailored to comparisons, Clutch provides the highest throughput among PuD approaches for vector–scalar comparisons. 
Second, Clutch takes an algorithmic approach rather than an architectural one, and therefore can be applied to modern density-optimized DRAM chips~\cite{mutlu2013memory, nam2024dramscope, marazzi2024hifi} without requiring modifications to DRAM circuitry.

Among bit-serial PuD systems, recent successors to SIMDRAM, such as MIMDRAM~\cite{oliveira2024mimdram} and Proteus~\cite{oliveira2025proteus}, add modifications to DRAM circuitry or its peripheral logic to support MIMD execution and dynamic bit-precision, respectively. Clutch is orthogonal to these architectural enhancements and compatible with such PuD architectures.

Several non-bit-serial PuD techniques have also been explored, including designs that support inter-column data movement~\cite{li2017drisa, li2018scope, deng2023dram}, stochastic computing~\cite{li2018scope,thakkar2023low,afifi2024artemis,shivanandamurthy2021atria}, and LUT-based execution~\cite{deng2019lacc,sutradhar2020ppim, zhou2022red, zhou2022lt, ferreira2022pluto}. \dtcr{3}{These techniques differ from Clutch in two important ways.} First, they require modifications to the DRAM cell array or its peripheral circuitry, increasing manufacturing cost and complexity, whereas Clutch takes an algorithmic approach that can operate on Unmodified DRAM~\cite{kubo2025mvdram, kubo2025pudtune, jahshan2024majork, garzon2026cadm, gao2022fracdram, olgun2022pidram, yuksel2024functionally, yuksel2023pulsar, yuksel2024simultaneous}. \dtcr{3}{Second, their designs primarily target multiplication and vector inner products rather than comparison operations. Among these, pLUTo~\cite{ferreira2022pluto} is the most closely related to Clutch as it enables general-purpose LUT-based function evaluation in DRAM.  However, in pLUTo's scheme, the memory footprint grows exponentially with operand bit-width when applied to vector–scalar comparison. Clutch addresses this limitation through its divide-and-conquer chunking strategy, which enables memory-efficient scaling to high bit-precisions (e.g., 16-bit and 32-bit).}

\subsubsection*{Processing-near-Memory (PnM)}
Processing-near-memory (PnM) adds custom logic near the memory array~\cite{ahn2015scalable, ahn2015pim, akin2015datareorg, akin2014hamlet, alves2016largevector, alves2015vectorops, alves2015savingmovements, asghari2016singleisa, asghari2016chameleon, azarkhish2016logicbase, azarkhish2016case, azarkhish2018neurostream, babarinsa2015jafar, besta2021sisa, boroumand2020practical, boroumand2021googleedge, boroumand2021mitigatingedge, boroumand2018googleworkloads, boroumand2017lazypimcal, boroumand2021polynesiaarxiv, boroumand2022polynesia, boroumand2019conda, boroumand2017lazypimarxiv, cali2020genasm, cho2020mcdramv2, dai2018graphh, delima2018designspace, denzler2021casper, devaux2019truepim, drumond2017mondrian, elliott1999computationalram, farmahini2014drama, farmahini2015nda, fernandez2020natsa, gao2016hrl, gao2017tetris, ghiasi2022genstore, giannoula2022sparsep, giannoula2021syncron, gokhale1995terasys, gomezluna2021benchmarkingcut, gomezluna2022benchmarkingaccess, gomezluna2021benchmarkingarxiv, gu2016hardwaresecurity, gu2020ipim, guo2014memoryside, hadidi2017cairo, hall1999diva, hsieh2016tom, hsieh2016pointerchasing, huang2019activerouting, huang2020heterogeneouspim, jang2019charon, ke2021axdimm, kersey2017lightweightsimt, kim2016neurocube, kim2018grimfilter, kim2017grimfilterarxiv, kwon2021fimdram, lee2015bssync, lee2021pimproduct, lee2022gddr6aim, li2019pims, lim2017tep, liu2018heterogeneoustraining, lockerman2020livia, matam2019graphssd, nai2017graphpim, nair2015activememorycube, niu2022processnearmemory, oliveira2017nim, patterson1997intelligentram, pattnaik2016gpupim, pugsley2014ndc, santos2018processing3d, santos2017operandsize, shin2018mcdram, singh2019napel, singh2020nero, sun2021abcdimm, tsai2018adaptivescheduling, xi2015beyondwall, zhang2014top, zhang2018graphp, zhu2013spgemm3d, zhuo2019graphq}. PnM exploits the high internal memory bandwidth \dtcr{3}{inside a DRAM chip}.
While PnM accelerates computation by \dtcr{3}{providing higher memory bandwidth and lower latency} to dedicated processing units, the full input vector must still be transferred from the memory array to those units.
In contrast, PuD performs comparisons directly \emph{within} the memory array, \dtcr{3}{avoiding the transfer of the input vector to logic units}.
Furthermore, PuD requires no additional processing logic and directly exploits the massive column-level parallelism of the existing DRAM arrays.

\section{Conclusion}
We present Clutch, a PuD-oriented algorithm \dtcr{3}{and data representation} for accelerating vector–scalar comparison. \dtcr{3}{Clutch provides a flexible tradeoff between memory footprint and throughput by combining a lookup-table-based comparison using temporal coding with a divide-and-conquer approach.}  Across two applications, GBDT inference and predicate evaluation, Clutch improves end-to-end throughput (and energy efficiency) by an average of $12\times$ ($69\times$) over optimized processor execution and $2.9\times$ ($3.0\times$) over the state-of-the-art bit-serial PuD approach. \dtcr{3}{These results demonstrate that DRAM can serve as a high-performance and energy-efficient computing substrate for comparison-intensive workloads.}

\begin{acks}
\dtcr{2}{We thank the anonymous reviewers of MICRO 2025, HPCA 2026, and ICS 2026 for their valuable feedback. We also thank the members of CASYS at UTokyo and the SAFARI Research Group at ETH Zurich for providing a stimulating intellectual environment.
We acknowledge the generous gifts from our industrial partners, including Google, Huawei, Intel, and Microsoft. 
This work is supported in part by JST CREST (JPMJCR21D2), JSPS KAKENHI (23H00467), JST ACT-X (JPMJAX25CC), the Semiconductor Research Corporation (SRC), the ETH Future Computing Laboratory (EFCL), and the AI Chip Center for Emerging Smart Systems (ACCESS).}
\end{acks}

\bibliographystyle{ACM-Reference-Format}

\end{document}